\documentclass[fleqn,usenatbib]{mnras}
\usepackage{amsmath}
\usepackage{graphicx}
\usepackage{epstopdf}
\usepackage{amssymb,txfonts}
\title[Hot flows in NGC 4258 and NGC 7213]{Hot accretion flows in low-luminosity active galactic nuclei in NGC 4258 and NGC 7213}
\author[M. Szanecki et al.]
{Micha{\l} Szanecki$^1$, Andrzej Nied{\'z}wiecki$^1$ and Rafa{\l} Wojaczy{\'n}ski$^2$\\
$^1$Faculty of Physics and Applied Informatics, {\L}{\'o}d{\'z} University, Pomorska 149/153, 90-236 {\L}{\'o}d{\'z}, Poland\\
$^2$AstroCeNT, Nicolaus Copernicus Astronomical Center of the Polish Academy of Sciences, 00-614 Warsaw, Poland\\
}

\pagerange{\pageref{firstpage}--\pageref{lastpage}}
\pubyear{2022}

\begin{document}
\maketitle

\label{firstpage}

\begin{abstract}
We study the high energy emission of two active galactic nuclei (AGN), NGC 4258 and  NGC 7213. We directly apply the general-relativistic (GR) hot flow model, {\tt kerrflow}, to the archival {\it BeppoSAX}, {\it NuSTAR} and {\it Suzaku} observations of these objects. Most of these data sets indicate that about 10--20 per cent of the accretion power is used for the direct heating of electrons, however,  we find also indications for significant changes of the electron heating efficiency in some cases. Furthermore, all these X-ray data sets indicate rather strongly magnetized flows, with the magnetic field close to the equipartition with the gas pressure. Comparison of the model prediction with the {\it Fermi}/LAT data for NGC 7213 allows us to constrain the content of nonthermal protons to at most 10 per cent. 
\end{abstract}

\section{Introduction}
\label{sec:intro}

Optically thin, hot accretion flows (a.k.a.\ advection dominated or radiatively inefficient accretion flows) represent a widely accepted description of black holes accreting at low accretion rates. This hot mode of accretion has a strong physical motivation  \citep[e.g.][]{1976ApJ...204..187S,1977ApJ...214..840I,1995ApJ...452..710N} and shows a good agreement with low-luminosity black-hole systems, in particular with nearby active galactic nuclei \citep[e.g.][]{2008ARA&A..46..475H,2014ARA&A..52..529Y}. Spectral models  of the hot flows have been applied to constrain the accretion rate and other parameters of the accretion flows of these low-luminosity AGNs in several works \citep[e.g.][]{1996ApJ...462..142L,1997ApJ...489..791M,2009ApJ...699..513L,2014MNRAS.438.2804N,2016MNRAS.463.2287X}. However, the applied models involved several approximations making the accuracy of these comparisons somewhat uncertain. The crudest of these concerns non-relativistic treatments of thermal Comptonization, which fail at relativistic temperatures, typically found in hot flow solutions, and give very inaccurate X-ray spectra \citep[see figure 5 in][]{2022ApJ...931..167N}. Further shortcomings include a non-GR description of both hydrodynamics and radiative transfer, local approximations of Comptonization, and the neglect of nonthermal particles.
Aiming at the improvement of these weaknesses, we developed a precise, fully GR spectral model of hot flows, {\tt kerrflow} \citep{2022ApJ...931..167N}. Here we apply it to two nearby, intrinsically faint LINER/Seyfert 1 galaxies, NGC 4258 and NGC 7213. 

NGC 4258 ($z=0.00149$), radiating at $L/L_{\rm Edd} \sim 10^{-4}$ \citep{2009ApJ...691.1159R}, is an excellent source for testing the accretion models, owing to the precise measurement of its supermassive  black hole mass and distance \citep{1995Natur.373..127M,2013ApJ...775...13H,2019ApJ...886L..27R} and was among objects whose observed properties supported the development  of the advection-dominated solution \citep{1996ApJ...462..142L}.

NGC 7213 ($z = 0.00584$), with $L/L_{\rm Edd} \sim 10^{-3}$ \citep{2005MNRAS.356..727S,2012MNRAS.424.1327E}. 
 does not show spectral indications for an inner, optically thick disc \citep{2005MNRAS.356..727S,2003A&A...407L..21B,2010MNRAS.408..551L} and has been described by the hot-flow model e.g. by \citet{2016MNRAS.463.2287X}. 
Its supermassive  black hole mass is relatively poorly estimated to $M = 8_{-6}^{+16} \times 10^7$ M$_\odot$ from 
the stellar velocity dispersion \citep{2014MNRAS.438.3322S} and $5 \times 10^7 < M/M_\odot < 2 \times 10^8$ from the reverberation of the H$\alpha$ line \citep{2017MNRAS.472.2170S}.
Nevertheless, this source has a major advantage for our study as its intrinsic X-ray emission is negligibly affected by internal absorption or reflection \citep[e.g.][]{2003A&A...407L..21B,2010MNRAS.408..551L}.

We focus our study on the X-ray emission, which is supposed to be produced in the innermost parts of an accretion flow. The current quality of X-ray observations of nearby AGNs does not allow us to reliably estimate the hot-flow parameters by the analysis of a single spectrum measured from a given object, due to degeneracies between the model parameters, see \citet{2022ApJ...931..167N}. 
These degeneracies are broken and the parameter values can then be reliably constrained when several spectra are analysed jointly. 
The optical depth of hot flows with low accretion rates is small, which implies a low efficiency of Compton cooling and hence a large electron temperature. Then, the Comptonization spectra produced at these low accretion rates exhibit pronounced Compton scattering bumps, which property can be used to verify the applicability of the hot-flow model and constrain the range of parameters.
In this work we use  the archival data of {\it BeppoSAX}, {\it NuSTAR} and {\it Suzaku} observations of NGC 4258 and NGC 7213. Observations by these satellites were chosen because they provide a wide energy coverage, which is crucial for the investigation of departures of an intrinsic spectrum from a simple power-law shape, predicted by the hot flow model at low accretion rates.

We also consider the $\gamma$-ray emission, which can be expected in hot flows because protons in their inner parts have energies above the threshold for pion production \citep{1997ApJ...486..268M,2003MNRAS.340..543O,2013MNRAS.432.1576N}. We compare the $\gamma$-ray fluxes predicted for parameters estimated from the analysis of the X-ray spectra with the {\it Fermi}/LAT observations of NGC 4258 and NGC 7213 and we find that for the latter object this gives some interesting constraints on the acceleration efficiency.

\section{Data reduction}
\label{sec:data}

\begin{table}
\caption{Observations studied in this work. For {\it BeppoSAX} the exposure time is given for LECS/MECS/PDS.}
\begin{center}
%\resizebox{\columnwidth}{!}{%
\begin{tabular}{cccccc}
 \hline
 Dataset & Obs.~ID & Satellite  & Exp.~[ks] & Ref.\\
 \hline
 \multicolumn{5}{c}{NGC 4258} \\
 \hline
 S1 & 701095010 & {\it Suzaku} & 100 & 1 \\
 \hline
 S2 & 705051010 & {\it Suzaku} & 104 & 2 \\
 \hline
 N1 & 60101046002 & {\it NuSTAR} & 55 & 3 \\
 \hline
 N2 & 60101046004 & {\it NuSTAR} & 104 & 3 \\
 \hline
 B & 50491001 & {\it BeppoSAX} & 33/99/47 & 4  \\
\hline
 \multicolumn{5}{c}{NGC 7213} \\
 \hline
 S & 701029010 & {\it Suzaku} & 91 & 5   \\
 \hline
 N & 60001031002 & {\it NuSTAR} & 102 & 6  \\
 \hline
 B1 & 50801002 & {\it BeppoSAX} & 59/48/56 & 7 \\
 \hline
 B2 & 508010021 & {\it BeppoSAX} & 15/15/34 & 7 \\
 \hline
 B3 & 508010022 & {\it BeppoSAX} & 20/46/31 & 7 \\
 \hline
 B4 & 51141002  & {\it BeppoSAX} & 40/61/38 & 7 \\
 \hline 
\end{tabular}\\
%}\\
{\it References}
1: \citet{2009ApJ...691.1159R}, 2: \citet{2016ApJ...831...37K}, 3: \citet{2022A&A...663A..87M}, 4: \citet{2001ApJ...556..150F}, 5: \citet{2010MNRAS.408..551L}, 6: \citet{2015MNRAS.452.3266U}, 7: \citet{2003A&A...407L..21B} 
\end{center}
\label{tab:data}
\end{table}

In the X-ray range we consider data from the XIS and PIN detectors on-board {\it Suzaku}, FPMA and FPMB on-board {\it NuSTAR} and LECS, MECS and PDS on-board {\it BeppoSAX}. 
For NGC 4258, we use the two {\it Suzaku} observations performed on 10-Jun-2006 and 11-Nov-2010, two {\it NuSTAR} observations performed on 16-Nov-2015 and 10-Jan-2016 and the {\it BeppoSAX} observation on 19-Dec-1998, referred to, respectively, as spectrum S1, S2, N1, N2 and B. 
For NGC 7213 we use the {\it Suzaku} observation on 22-Oct-2006, the {\it NuSTAR} observation on 5-Oct-2014 and the four {\it BeppoSAX} observations performed between 1999 and 2001, referred to, respectively, as spectrum S, N and B1 -- B4.
The X-ray spectral sets are summarized in Table \ref{tab:data}.

For {\it BeppoSAX} we use the ready, reprocessed data products available from the archive site\footnote{https://heasarc.gsfc.nasa.gov/FTP/sax}. 
%Background and response files were taken from \footnote{ftp://heasarc.gsfc.nasa.gov/sax/cal/bgd/} and \footnote{ftp://heasarc.gsfc.nasa.gov/sax/cal/responses/} respectively.  % Not sure if all links needed - RW
In the reduction of the {\it Suzaku} and {\it NuSTAR} data for both sources we used the standard procedures.

For {\it Suzaku}, event files from version 2.0.6.13 of the pipeline processing were used. Event files were reprocessed and screened using {\tt aepipeline} ftool v1.1.0 with default parameters and CALDB files from version v2015.03.12. Source spectra from the XIS CCDs were extracted from circular regions centred on the source, in the on-axis XIS nominal pointing position. The radius of the extraction region was set to 3.25 arcmin for NGC 4258 and 2.6 arcmin for NGC 7213. The background spectra were extracted from the circular regions with the same sizes, offset from the source region to avoid the sources on the corners of the CCD chips. XIS response files and ancillary response files were
generated using the XISRMFGEN and XISSIMARFGEN ftools, respectively.
Spectra and responses from the three front-illuminated XIS chips 0, 2 and 3 were added together using {\tt addascaspec} ftool. The resulting combined spectra are referred to as XISFI. The spectrum from the back-illuminated XIS 1 chip is referred to as XISBI. 

The {\it NuSTAR} data were reduced with the standard pipeline in the {\it NuSTAR} Data Analysis Software v1.8.0, using calibration files from CALDB v20191219. Spectra and light curves were extracted from the cleaned event files using the standard tool NUPRODUCTS.
The source data were extracted
from circular regions with the radius of 75 arcsec for NGC 7213 and 30 arcsec for NGC 4258. The background were extracted from blank areas close to the sources.

In the spectral analysis the 1.7-10 keV band for the MECS, the 0.2-4 keV band for the LECS and 15-200 keV band for PDS  were used. 
For the FPM detectors we use the 3--70 keV range and for PIN 15-50 keV range. For XIS we use the 0.3--10 keV keV range,  neglecting the data in the 1.7--1.9 keV range due to the known uncertainties in calibration associated with the instrumental Si K edge.
The spectra of all X-ray instruments were rebinned with the condition of at least 25 counts per bin.
The considered observations were previously studied in the works referenced in Table \ref{tab:data},  and we found our data of all detectors to be very similar to those presented in these works. 

Our main results were obtained by fitting jointly the B1-B4, N and S spectra for NGC 7213 and N1, N2, S1, S2 and B spectra for NGC 4258. To account for cross-calibration differences
between the considered detectors, we adopted multiplicative cross-normalization constants, with constraints reported by \citet{2017AJ....153....2M} and \citet{2015ApJ...812..116B}, respectively, on relative {\it Suzaku}/{\it NuSTAR} and {\it Beppo}/{\it NuSTAR} normalization; the count rates for PKS 2155-304 and NGC 1068, studied in these works, are comparable to those in NGC 4258 and NGC 7213. We also took into account the ranges of internal cross-calibration for {\it BeppoSAX} detectors suggested by the {\it BeppoSAX Cookbook}\footnote{https://heasarc.gsfc.nasa.gov/docs/sax/abc/saxabc} and for {\it Suzaku} given in the {\it Suzaku Technical Description}\footnote{https://heasarc.gsfc.nasa.gov/docs/suzaku/prop\_tools/suzaku\_td/suzaku\_td.html}.
We then normalized all detectors relative to FPMA and allowed the relative normalizations to vary in the following ranges: 0.6-0.8 for PDS, 0.98-1.04 for MECS, 0.69-1.04 for LECS, 0.91-0.97 for XIS/FI and XIS/BI. The FPMB normalization was allowed to fit freely but in all cases we found it to be consistent within 2 per cent with FPMA. Also XIS/BI is consistent in all cases with XIS/FI within 2 per cent. The PIN normalization relative to XIS/FI was fixed at 1.16 (we checked that allowing it to vary in the range 1.11--1.21 does not affect any of our results).

For the $\gamma$-ray emission we performed a maximum likelihood analysis
using the Pass 8 LAT data in the 0.1--100 GeV range. We analysed data between 2008 August 4 and 2020 February 6. We used {\tt Fermitools 1.2.1} with the P8R3\_SOURCE\_V2  response
function. In the model we used standard templates for the Galactic (gll\_iem\_v07.fits) and the isotropic (iso\_P8R3\_SOURCE\_V2\_v1.txt)
backgrounds. We took into account all sources reported in the 8-year LAT
Catalog \citep{2020ApJS..247...33A} within the radius of $10^\circ$  around each object.

\section{Spectral analysis}
\label{sec:results}

\begin{figure}
\centering
 \includegraphics[width=8cm]{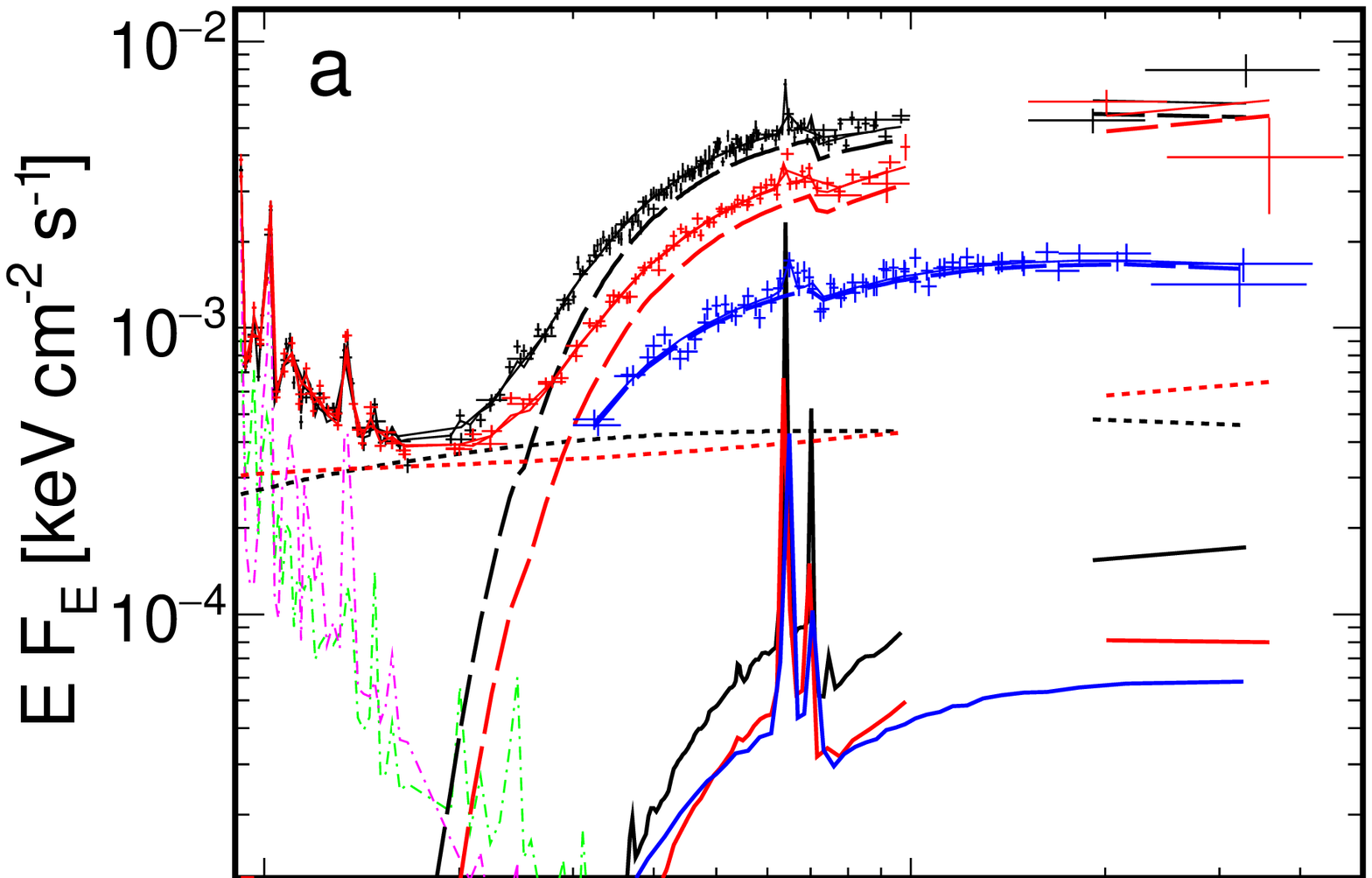}
 \includegraphics[width=8cm]{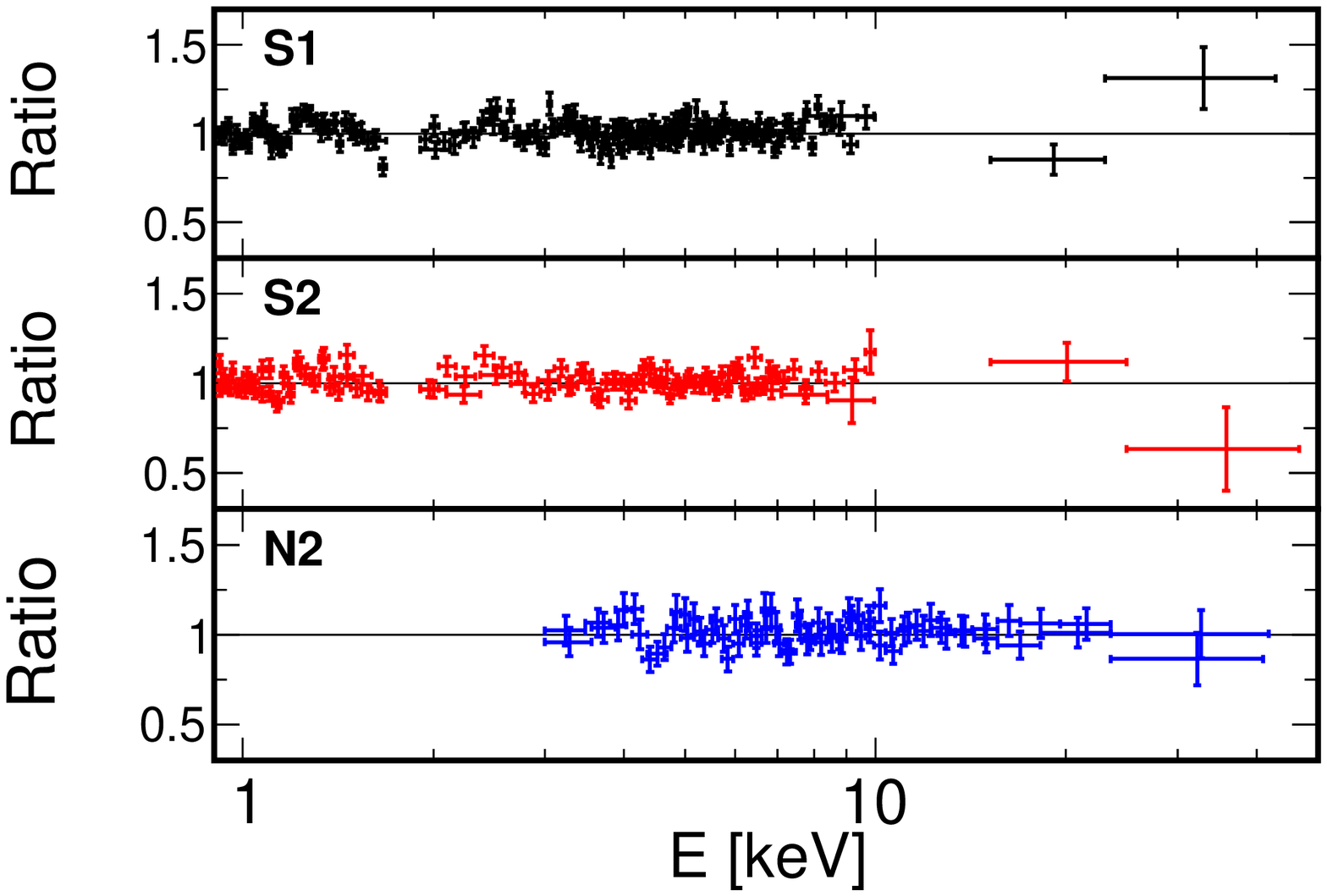}
\caption{Top panel: the unfolded data and model components for our best-fitted model with free $\delta$ to NGC 4258, see Table \ref{tab:fit_4258}. For clarity we show only spectrum S1 (black), S2 (red) and N2 (blue). The absorbed {\tt kerrflow} spectra are shown by the dashed lines. The  \texttt{borus} spectra are shown by the solid lines.  The unabsorbed contributions of {\tt kerrflow} are shown by the dotted lines. The two \texttt{apec} components are shown by the green and magenta dot-dashed lines. 
The data were rebinned for clarity of the figure. Bottom panels: the data to model ratios. }
\label{fig:spectra4258}
\end{figure}

\subsection{Spectral components}

We briefly describe here the models used in our spectral analysis.

The hot-flow model, {\tt kerrflow}, is parametrized by the black hole mass, $M$, and spin parameter, $a$, the accretion rate parameter, $\dot m = \dot M / \dot M_{\rm Edd}$ (where $\dot M_{\rm Edd}= L_{\rm Edd}/c^2$), the plasma magnetization, $\beta$, defined as the ratio of the gas pressure to the magnetic pressure, and the electron heating efficiency $\delta$, defined as the fraction of the dissipated energy which heats directly electrons. The model normalization, $N$, is determined by $M$ and the distance, $d$. 
Only the seed photons produced internally within the flow by bremsstrahlung (negligible at the fitted values of $\dot m$) and synchrotron emission are taken into account. The hybrid energy distribution of electrons is taken into account with the nonthermal component from the decay of charged pions,  whose production must occur in hot flows due to the presence of energetic protons \citep{1999MNRAS.304..501M,2015ApJ...799..217N}. In the range of luminosities studied here, the production of X-rays is fully dominated by thermal Comptonization. The predicted X-ray spectra exhibit a significant dependence on the model parameters, but the current quality of X-ray data from nearby AGNs appears inadequate for reliable estimation of the parameter values, especially if $M$ (and then the normalization) is poorly constrained. 
However, fitting simultaneously several X-ray spectra of an object allows to test the model in more detail, as the change of one or more model parameters must now reproduce the observed spectral variations, and to robustly constrain the parameters; 
see figures 3 and 4 and discussion in \citet{2022ApJ...931..167N}.

Both NGC 4258 and NGC 7213 show signatures of a diffuse, collisionally excited emission in  the soft X-ray range. 
We described it with the \texttt{apec} model\footnote{http://atomdb.org}. This emission should not exhibit significant variability, therefore, we included it for all datasets with linked parameters, i.e.\ temperature and normalization. 

Also, both AGNs exhibit the Fe K$\alpha$ line at $6.4$ keV, indicating reprocessing of the X-ray radiation in a neutral matter. We used for it the {\tt borus}  model \citep{2018ApJ...854...42B}, which self-consistently computes the intensity of the iron fluorescent lines and the shape of the reprocessed continuum. This allows us to check if the Compton continuum associated with the iron lines affects our fitting results. We used the version {\tt borus02} parametrized by the relative iron abundance, $Z_{\rm Fe}$, the column density, $N_{\rm H}$ (spanning a range below and above the Compton-thick threshold), the inclination angle, $i$, the covering factor, $f_{\rm cov}$, 
and the parameters of the incident spectrum, assumed to be a power-law with the photon spectral index, $\Gamma$, with an exponential cut-off at $E_{\rm cut}$. In our spectral models assuming that the incident spectrum is formed in the hot flow, we used $\Gamma$ and $E_{\rm cut}$ giving the closest match for the fitted {\tt kerrflow} spectra.
We also attempted to describe the neutral reprocessing using the  {\tt hreflect} model \citep{2019MNRAS.485.2942N}, 
which is a convolution model allowing us to take into account the incident spectrum in the actual form given by {\tt kerrflow}. However, we did not find improvements compared to the model with {\tt borus}.
Finally, we attempted to smear the {\tt borus} spectrum using the {\tt rdblur} model
to check if the Doppler broadening or shift of this component improves the spectral description, however, we did not find any indications for it in either object.

For the internal absorption, we used the Teubingen-Boulder absorption model, {\tt ztbabs}.

\begin{table*}
 \caption{Parameters of our best spectral model fitted jointly to the five data sets of NGC 4258. We give the {\tt kerrflow} parameters for the two versions of the model, with a linked and unlinked $\delta$. The parameters of the remaining components are given for the model with a free $\delta$; for a linked $\delta$ they are similar.}
 \begin{tabular}{|l|c|c|c|c|c|}
   \hline
   \multicolumn{6}{|c|}{model: \texttt{TBabs*[apec$_{1}$+apec$_{2}$+cnst*kerrflow+zTBabs*(kerrflow+borus)]}} \\
   \hline
   & S1 & S2 & N1 & N2 & B \\
   \hline
   \texttt{apec$_{1}$} & \multicolumn{5}{|c|}{{\bf collisionally-ionized diffuse emission}} \\
   $kT_{\rm 1}$               & \multicolumn{5}{|c|}{$0.33^{+0.02}_{-0.02}$(l)}\\[0.1cm]
   ${\rm norm}_{\rm 1}$       & \multicolumn{5}{|c|}{$8^{+51}_{-2}\times 10^{-4}$(l)}\\[0.1cm]
   \texttt{apec$_{2}$}   &    &    &    &\\
   $kT_{\rm 2}$               & \multicolumn{5}{|c|}{$0.85^{+0.05}_{-0.05}$(l)}\\[0.1cm]
   ${\rm norm}_{\rm 2}$       & \multicolumn{5}{|c|}{$1.3^{+0.3}_{-0.3}\times 10^{-4}$(l)}\\[0.1cm]
   \hline
   			      & \multicolumn{5}{|c|}{{\bf scattering/leakedge}} \\
   $K$                            & $0.08^{+0.01}_{-0.01}$ & $0.12^{+0.01}_{-0.01}$ & - & - & $0.07^{+0.01}_{-0.01}$ \\[0.1cm]
   \hline
   \texttt{zTBabs} & \multicolumn{5}{|c|}{{\bf internal absorption}} \\
   $n_{\rm H}\; [10^{22}{\rm cm}^{-2}]$ & $10.6^{+0.3}_{-0.3}$ & $12.4^{+0.5}_{-0.4}$ & $6^{+1}_{-1}$ & $8^{+1}_{-1}$ & $10^{+1}_{-1}$ \\[0.1cm]
   \hline
    \texttt{borus02}  & \multicolumn{5}{|c|}{{\bf internal reprocessing}} \\
    $Z_{\rm Fe}$      & \multicolumn{4}{|c|}{$2^{+5}_{-1}$} & - \\[0.1cm]
    $f_{\rm cov}$  & $<0.5$ & $<0.4$ & $>0.3$ & $>0.1$ & - \\[0.1cm]
    \hline
    \texttt{kerrflow} & \multicolumn{5}{|c|}{{\bf hot flow, free $\delta$}} \\
    $a$      & \multicolumn{5}{|c|}{$0.4^{+0.2}_{-0.1}$(l)}\\[0.1cm]
    $\beta$  & \multicolumn{5}{|c|}{$1.0^{+0.2}_{-0}$(l)}\\[0.1cm]
    $\delta$ & $0.15^{+0.05}_{-0.12}$ & $0.001^{+0.001}_{-0}$ & $0.2^{+0.1}_{-0.2}$ & $0.11^{+0.02}_{-0.02}$ & $0.12^{+0.02}_{-0.02}$ \\[0.1cm]
    $\dot{m}\;[10^{-2}]$& $0.6^{+0.1}_{-0.1}$ & $3.3^{+3.8}_{-0.5}$ & $0.15^{+0.05}_{-0.03}$ & $0.22^{+0.06}_{-0.03}$ & $1.0^{+0.2}_{-0.2}$ \\[0.1cm]
    $N$  & \multicolumn{5}{|c|}{$0.95^{+0.10}_{-0}$(l)} \\[0.1cm]
    \hline

    $\chi^{2}/{\rm DoF}$ & \multicolumn{5}{|c|}{2036/1877}\\[0.1cm]
    $\chi^{2}$ & 653 & 658 & 324 & 310 & 91 \\[0.1cm]
    \hline
    \hline
    \texttt{kerrflow} & \multicolumn{5}{|c|}{{\bf hot flow, linked $\delta$}} \\
    $a$      & \multicolumn{5}{|c|}{$0.56^{+0.39}_{-0.15}$(l)}\\[0.1cm]
    $\beta$  & \multicolumn{5}{|c|}{$1.6^{+0.3}_{-0.2}$(l)}\\[0.1cm]
    $\delta$ & \multicolumn{5}{|c|}{$0.07^{+0.02}_{-0.02}$(l)}\\[0.1cm]
    $\dot{m}\;[10^{-2}]$& $1.2^{+0.1}_{-0.3}$ & $0.6^{+0.1}_{-0.2}$ & $0.30^{+0.03}_{-0.04}$ & $0.24^{+0.02}_{-0.02}$ & $1.2^{+0.2}_{-0.2}$ \\[0.1cm]
    $N$  & \multicolumn{5}{|c|}{$0.96^{+0.02}_{-0.01}$(l)} \\[0.1cm]
    \hline

    $\chi^{2}/{\rm DoF}$ & \multicolumn{5}{|c|}{2064/1881}\\[0.1cm]
    $\chi^{2}$ & 651 & 677 & 327 & 315 & 94 \\[0.1cm]
    \hline
  \end{tabular}\\
{\it Notes:}
Parameters denoted with '(l)' are linked across S1, S2, N1, N2 and B. The Galactic absorption is modelled using  \texttt{Tbabs} with $N_{\rm H} = 1.1 \times 10^{20}$ cm$^{-2}$. For {\tt kerrflow} $N=1$ corresponds to $M=3.98 M_\odot$ and $d = 7.58$ Mpc.
  \label{tab:fit_4258}
\end{table*}

\begin{figure}
\centering
 \includegraphics[width=8cm]{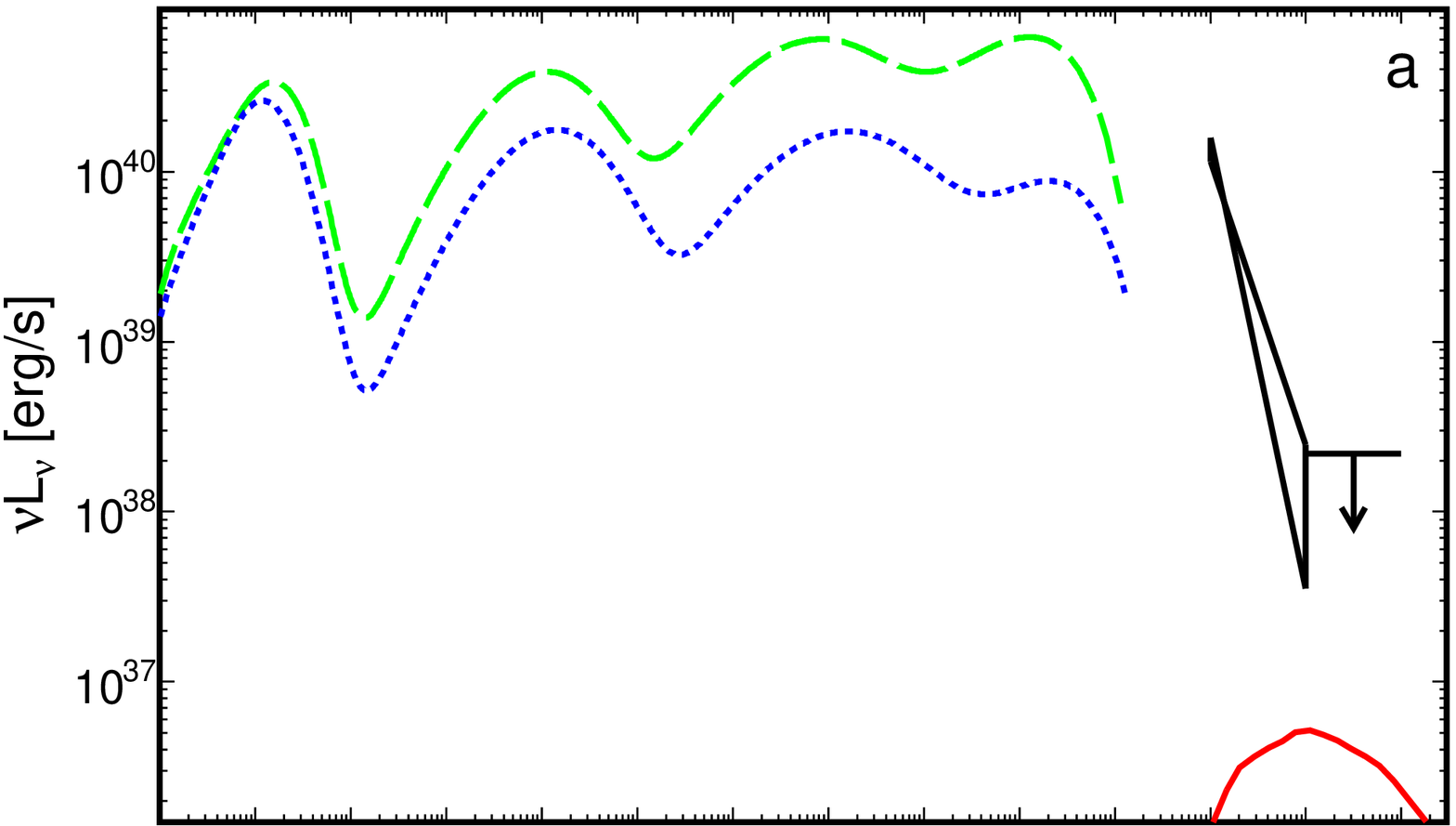}
 \includegraphics[width=8cm]{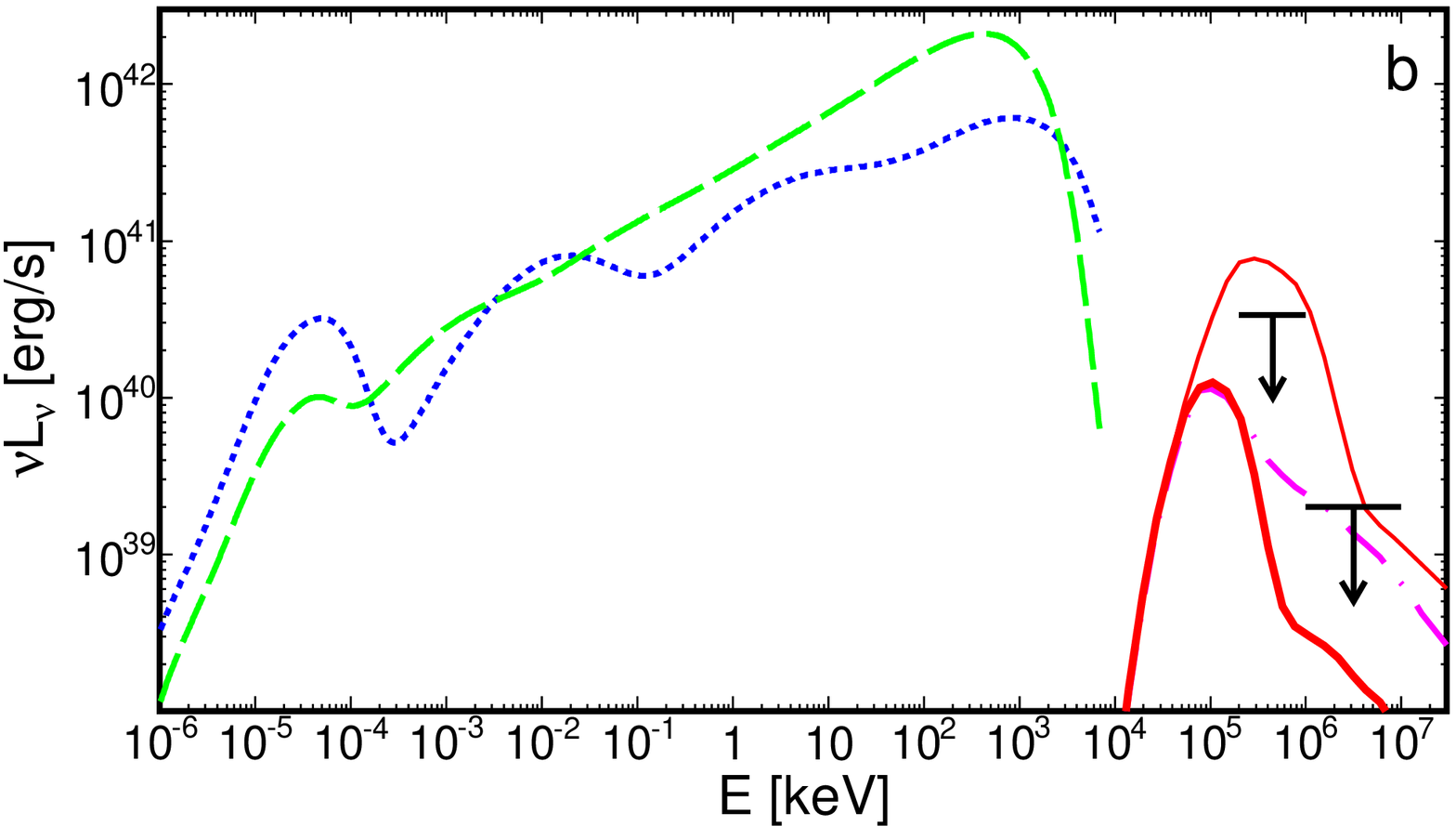}
\caption{Intrinsic spectral energy distributions of hot flows with parameters fitted to a) NGC 4258 and b) NGC 7213, and the results of the analysis of LAT data for the locations of these objects. a) The green dashed and dotted blue lines show the spectra of electron emission in the model with unlinked $\delta$ fitted to S1 and N2, respectively. The red solid line shows the observed spectrum of $\pi^0$-decay photons for $\dot m = 0.003$, $\delta=0.1$, $\beta=1$, $a=0.6$, $s=2.4$ and $\eta = 0.1$.
The region with the bow-tie shape shows parameters from the analysis of the $\sim 5 \sigma$ signal in 0.1--1 GeV range and the black arrow shows the upper limit for the 1--10 GeV range.
b) The green dashed and dotted blue lines show the spectra of electron emission in the model fitted to B2 and N, respectively.
The observed spectrum of $\pi^0$-decay photons for $\dot m = 0.1$, $\delta=0.1$, $\beta=3$, $a=0.95$ and $s=2.4$ is shown by the magenta dot-dashed line for $\eta=0.1$ and  the thicker red solid line for $\eta=0.01$.  The thinner red solid line shows the same spectrum for $\eta=0.01$ neglecting the internal $\gamma \gamma$ absorption.
The black arrows show the LAT upper limits for 0.2--1 GeV and 1--10 GeV ranges. 
}
\label{fig:sed}
\end{figure}

\begin{figure}
\centering
 \includegraphics[width=8cm]{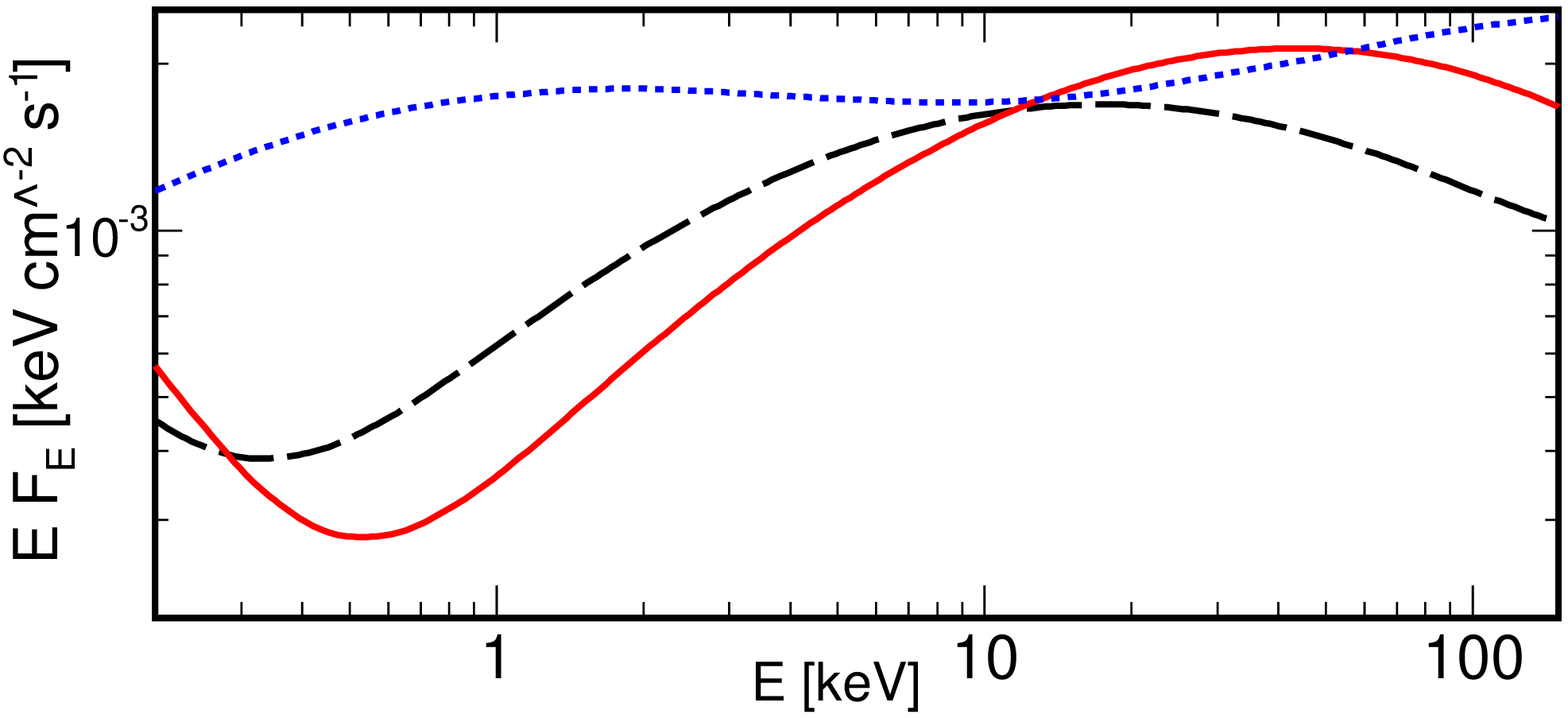}
\caption{Changes of the intrinsic {\tt kerrflow} spectra corresponding to the change of $\delta$ in the model for spectrum N2 of NGC 4258. The dashed black line shows the {\tt kerrflow} spectrum fitted for a free $\delta$, see Table \ref{tab:fit_4258}, i.e.\ with $\delta = 0.11$. The red solid and blue dotted lines show the {\tt kerrflow} spectra fitted with a fixed $\delta = 0.5$ and $10^{-3}$, respectively, see text.
}
\label{fig:delta}
\end{figure}

\subsection{NGC 4258}

The nuclear emission of NGC 4258 is strongly absorbed and the observed radiation is fully dominated by the diffuse emission 
at $\la 1$ keV. Therefore, in all detectors, we considered only the data above 0.9 keV. 

Guided by previous studies \citep{2009ApJ...691.1159R,2020ApJ...897...66L}, in our description  of the X-ray spectrum of NGC 4258 we included the  diffuse emission described by two {\tt apec} components (which gives a significantly better description, with $\Delta \chi^2 \simeq -200$, than a version with a single {\tt apec} component); the optically thin reprocessor described by {\tt borus02}; the internal absorber described by {\tt zTbabs}; the intrinsic emission described by {\tt kerrflow} and the additional component described by the unabsorbed {\tt kerrflow} (with parameters linked to the main component representing the intrinsic emission), attenuated by a factor $K \simeq 0.1$. The last component is very poorly constrained \citep[cf.][]{2009ApJ...691.1159R} and may be attributed e.g.\ to a scattering or leakage of the nuclear emission through a clumpy absorber, in which case $K$ would be interpreted as a scattering or leakage fraction. Furthermore, its contribution is needed only at $\la 3$ keV, therefore, we use it only for spectra S1, S2 and B.
Adding it to N1 and N2 did not affect the fitting results and for these two spectra $K$, was fitted at $\simeq 0$.
To reduce the number of free parameters, we linked the $N_{\rm H}$ of {\tt borus02} to the column density of {\tt zTbabs}.
The presence of the {\tt borus} component is significant 
in all spectra except for spectrum B. 
The definition of our spectral model is given in the header of Table \ref{tab:fit_4258}.

In our application of {\tt kerrflow} to NGC 4258 we normalized the model to $M=3.98 \times 10^7 M_\odot$ and $d = 7.58$
Mpc \citep{2019ApJ...886L..27R} and we allowed for the $\pm 5$ per cent uncertainty of the normalization, i.e.\ we considered the model normalization in the range $N = 1 \pm 0.05$.
We first linked $\delta$ and $\beta$ across spectra S1, S2, N1, N2 and B to check whether the observed spectral variations can be described by the change of a single parameter, $\dot m$. We found a good description with $\chi^2/{\rm DoF} = 2064/1881$ for  $\delta \simeq 0.07$ and $\beta \simeq 1$. We then allowed $\delta$ to vary, which slightly improved the fit by $\Delta \chi^2 = -28$ for 4 free parameters, with very low $\delta \simeq 0.001$ fitted to S2 and much larger $0.1 \la \delta \la 0.2$ fitted to the remaining spectra. We give parameters fitted in both versions (with either free or linked $\delta$) in Table \ref{tab:fit_4258}. Figure \ref{fig:spectra4258} shows the observed spectra S1, S2 and N2 and the fitted model with a free $\delta$. Allowing $\beta$ to vary did not improve the fit and for all spectra it was fitted at $\beta \simeq 1$.

In order to  check if our results depend on the assumed cross-normalization among the considered satellites, we also fitted jointly only the two {\it NuSTAR} and then only the two {\it Suzaku} spectra. Fitting jointly N1 and N2 we found the same parameters as given for these spectra in Table \ref{tab:fit_4258}. Fitting jointly S1 and S2 we found parameters differing by less than a factor of $\la 1.5$ from values given for these spectra in Table \ref{tab:fit_4258}. Then, in both cases, we recovered the results obtained with the full set of spectral data, which is primarily due to the tightly constrained normalization for this source.

Spectra produced at the fitted $\delta \simeq 0.1$ and $\dot m \la 0.01$ exhibit pronounced Compton scattering bumps and deviate from a power-law shape, see Figure \ref{fig:sed}a. In order to verify if the related spectral curvature
affects our spectral description, 
we also fitted our base model but with {\tt kerrflow} replaced by power-law spectra and we found $\chi^2/{\rm DoF} = 2040/1880$ for the joint fit (in this joint fitting only the parameters of \texttt{borus02} and \texttt{apec} where linked across S1, S2, N1, N2 and B). We note that although the model with {\tt kerrflow} formally has three more parameters than the one with power-law spectra, it actually has less freedom in fitting the observed data, due to spectral changes which are strictly related to the change of $\dot m$ or $\delta$ needed to explain the change of the flux. In contrast, the slopes and fluxes of the empirical power-law components are fitted completely freely.
The same goodness of the fits with {\tt kerrflow} and with the power-law spectra indicates 
that the quality of X-ray data from NGC 4258 does not allow us to confirm or rule out departures from a power law predicted at the fitted range of parameters. 
This is partially due to the spectral complexity of this source related to the strong internal absorption.
On the other hand, we can rule out stronger distortions from a power-law shape which are predicted at larger values of $\delta$. Setting $\delta=0.5$ we found $\chi^2/{\rm DoF} = 2191/1882$ 
and systematic residuals related to spectral bumps, which at this value of $\delta$  are more pronounced, as shown by the red solid line in Figure \ref{fig:delta}. We also rule out a constant $\delta = 10^{-3}$. Fixing this value of $\delta$ we found $\chi^2/{\rm DoF} = 2096/1882$ with systematic residuals which occur because the spectra at $\dot m > 0.01$ fitted in this case are too soft, see the blue dotted line in Figure \ref{fig:delta}.

NGC 4258 displayed a significant flare during the S1 observation and \citet{2009ApJ...691.1159R} and \citet{2009PASJ...61..309Y} noted some hints for spectral variability associated with this flare. We extracted the spectra of the high and low flux states following the \citet{2009PASJ...61..309Y} criterion, i.e.\ we divided the S1 observation into subsets with the 2--10 keV count rates higher and lower than 0.5 counts s$^{-1}$, respectively. The resulting spectra can be described, with no apparent residuals, using {\tt kerrflow} with all parameters fixed at the values found for S1 (as given in Table  \ref{tab:fit_4258} for the model with unlinked $\delta$), except for $\delta$ which is fitted at 0.13 and 0.16 for the higher and lower flux spectrum, respectively.

Finally, we note that although the spin parameter is formally determined at $a = 0.4^{+0.2}_{-0.1}$ in the joined fit with unlinked $\delta$, this parameter is rather poorly constrained, namely, setting $a=0$ yields $\Delta \chi^2 = +17$, whereas setting $a=0.95$ yields only $\Delta \chi^2 = +5$. This weak dependence on $a$ is due, in particular, to a strong contribution of the compressive heating of electrons at $\dot m \la 0.01$, which process shows a weak dependence on $a$ \citep[c.f.][]{2012MNRAS.420.1195N}. As a sanity check of the model, we repeated the fit reported in Table \ref{tab:fit_4258} but with $a$ unlinked across all datasets. We found that $a$ is essentially unconstrained in datasets S2, N1, N2 and B, with changes in the full range of $0 \le a \le 0.95$ leading to $|\Delta \chi^2| < 2$, and only in S1 we found a significant lower-limit with the spin constrained to $a \ga 0.3$.

\begin{figure}
\centering
 \includegraphics[width=8cm]{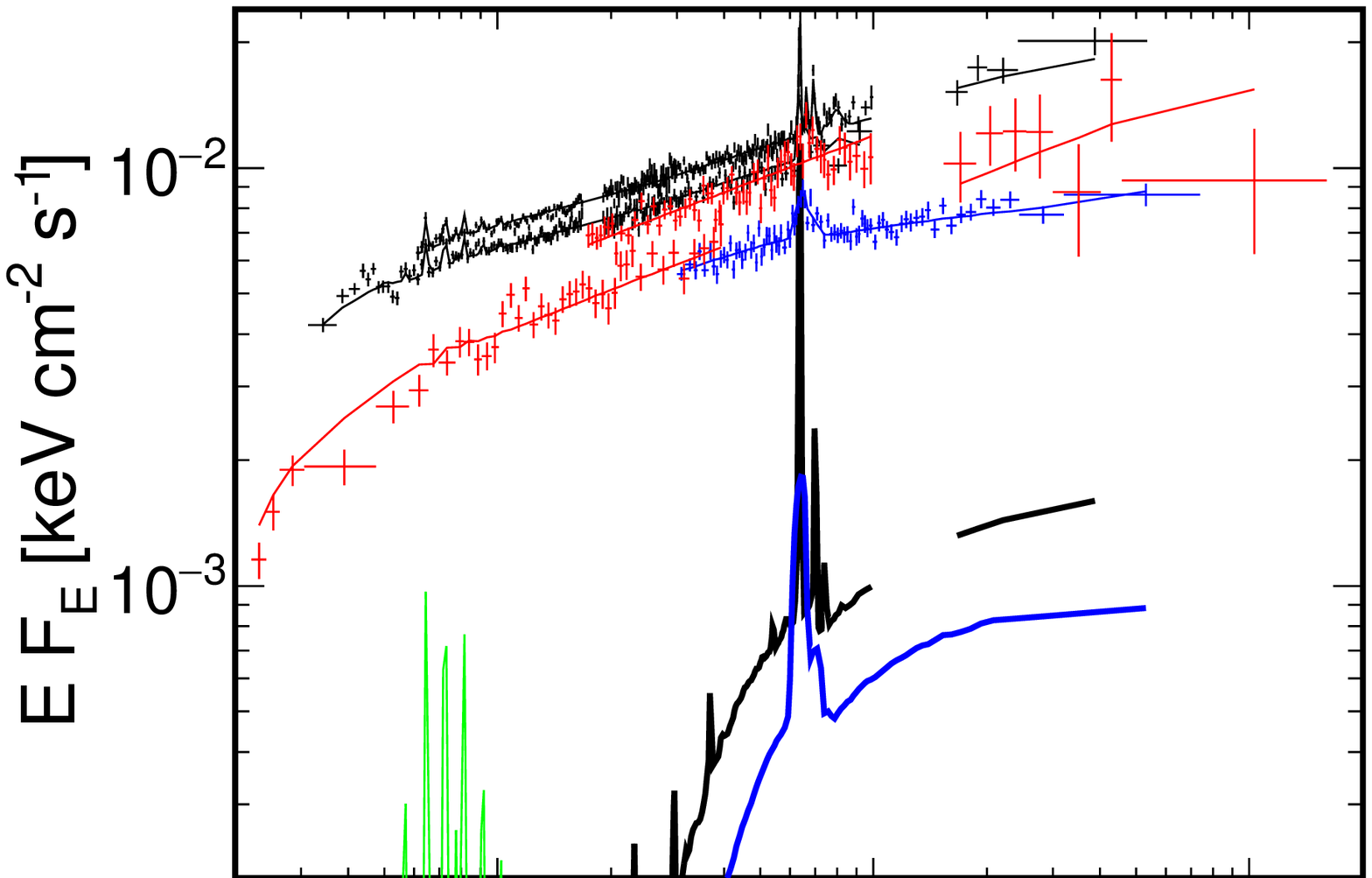}
 \includegraphics[width=8cm]{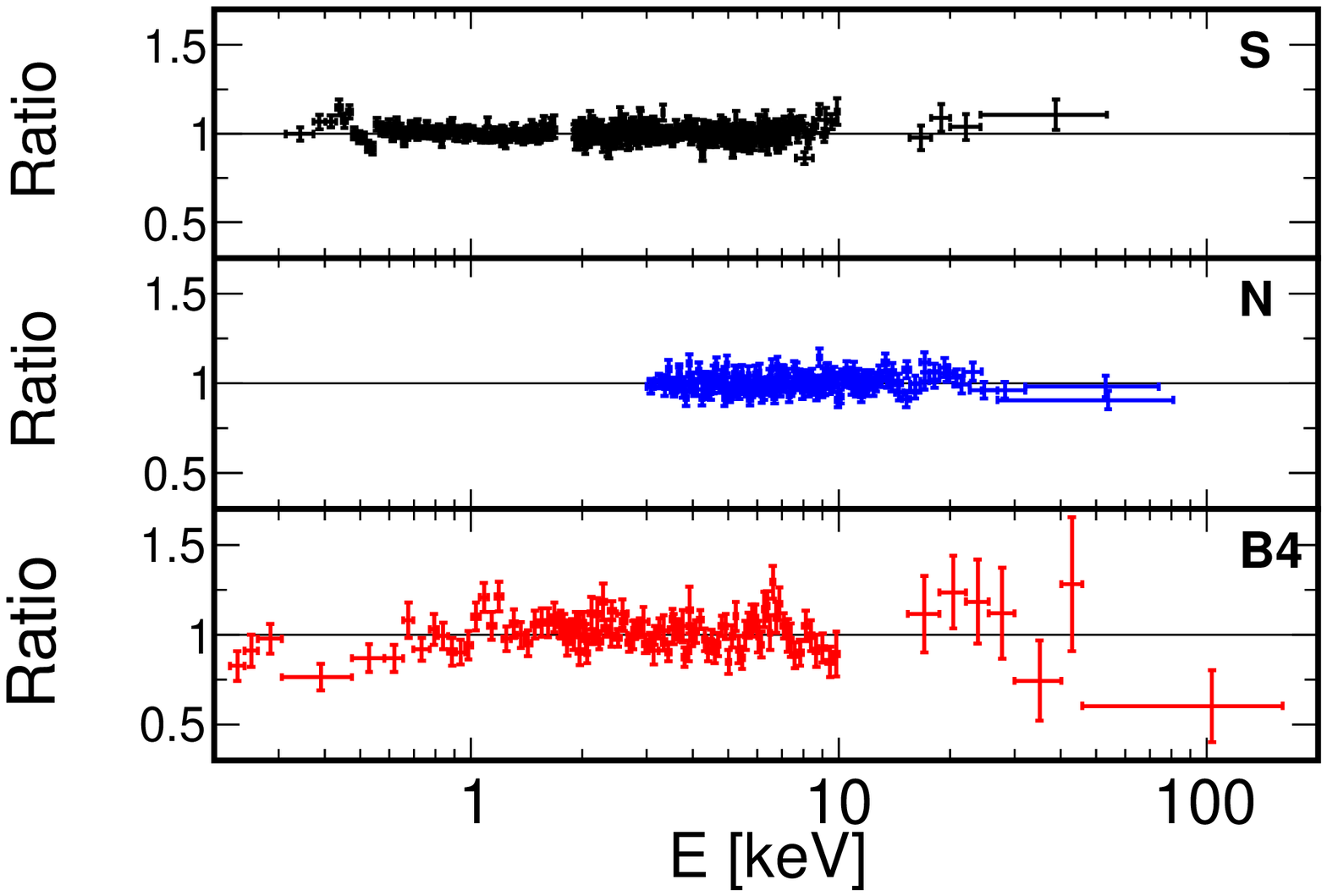}
\caption{Top panel: the unfolded data and model components for our best-fitted model to NGC 7213, see Table \ref{tab:fit}, for spectrum S (black), N (blue) and B4 (red). Spectra B1, B2 and B3 are similar to B4 and are not shown here for clarity. The thinner solid lines show the total model, the green line shows the \texttt{apec} spectrum, the thicker black and blue lines show the \texttt{borus} spectrum fitted in S and N, respectively. The data were rebinned for clarity of the figure. Bottom panels: the data to model ratios for the best-fit model for spectrum S, N and B4. }
\label{fig:spectra}
\end{figure}

\begin{table*}
 \caption{Parameters of our best spectral model fitted jointly to the six data sets of NGC 7213. The bottom row gives the contribution of each spectral set to the total $\chi^{2}$.}
 \begin{tabular}{|l|c|c|c|c|c|c|}
   \hline
   \multicolumn{7}{|c|}{model: \texttt{TBabs*(apec+adaf+rdblur*borus+zgauss+zgauss+zgauss)}} \\
   \hline
   & S & N & B1 & B2 & B3 & B4 \\
   \hline
   \texttt{apec} & \multicolumn{6}{|c|}{{\bf collisionally-ionized diffuse emission}} \\
   $kT$      & \multicolumn{6}{|c|}{$0.31^{+0.05}_{-0.03}$(l)}\\[0.1cm]
   norm            & \multicolumn{6}{|c|}{$1.2^{+0.3}_{-0.3}\times 10^{-4}$(l)}\\[0.1cm]
    \hline
    \texttt{borus}  & \multicolumn{6}{|c|}{{\bf internal reprocessing}} \\
    $Z_{\rm Fe}$        & \multicolumn{2}{|c|}{$1.0^{+0.1}_{-0.2}$} & - & - & - & -\\[0.1cm]
	$n_{\rm H}\; [10^{22}{\rm cm}^{-2}]$  & $8^{+1}_{-1}$ & $13^{+3}_{-3}$ & - & - & - & -\\[0.1cm]
%    $\log_{10}(n_{\rm H})$  & $22.9^{+0.03}_{-0.08}$ & $23.1^{+0.1}_{-0.1}$ & - & - & - & -\\[0.1cm]
    $f_{\rm cov}$  & $1^{+0}_{-0.9p}$ & $1^{+0}_{-0.8}$ & - & - & - & -\\[0.1cm]
    \hline
    \texttt{zgauss} & \multicolumn{6}{|c|}{{\bf Fe XXV line}} \\
    $E\;[\rm keV]$ & $6.67^{+0.03}_{-0.04}$ & $6.72^{+0.02}_{-0.06}$ & - & - & - & -\\[0.1cm]
    $\sigma\;[\rm eV]$ & $<87$ & $249^{+164}_{-133}$ & - & - & - & -\\[0.1cm]
    N [$10^{-6}$ph/(cm$^2$s)]& $7.4^{+4.4}_{-2.4}$ & $10.7^{+3.3}_{-4.4}$ & - & - & - & -\\[0.1cm]
    \hline
    \texttt{zgauss} & \multicolumn{6}{|c|}{{\bf  Fe XXVI line}} \\
    $E\;[\rm keV]$ & $6.95^{+0.03}_{-0.04}$ & $6.99^{+0}_{-0.08}$ & - & - & - & -\\[0.1cm]
    $\sigma\;[\rm eV]$ & $<110$ & $<270$ & - & - & - & -\\[0.1cm]
    N [$10^{-6}$ph/(cm$^2$s)]& $7.5^{+1.9}_{-2.1}$ & $6.2^{+2.3}_{-4.6}$ & - & - & - & -\\[0.1cm]
    \hline
    \texttt{zgauss} & \multicolumn{6}{|c|}{{\bf 8 keV line}} \\
    $E\;[\rm keV]$ & $8.0$(f) &- & - & - & - & -\\[0.1cm]
    $\sigma\;[\rm eV]$ & $250$(f) & - & - & - & - & -\\[0.1cm]
    N [$10^{-6}$ph/(cm$^2$s)]& $11.0^{+3.5}_{-3.5}$ & - & - & - & - & -\\[0.1cm]
    \hline
    \texttt{adaf} & \multicolumn{6}{|c|}{{\bf hot flow}} \\
    $a$      & \multicolumn{6}{|c|}{$0.95^{+0}_{-0.03}$(l)}\\[0.1cm]
    $\beta$  & \multicolumn{6}{|c|}{$2.9^{+0.1}_{-0.2}$(l)}\\[0.1cm]
    $\delta$ & $0.005^{+0.001}_{-0.001}$ & $0.50^{+0}_{-0.03}$ & $0.19^{+0.01}_{-0.02}$ & $0.16^{+0.06}_{-0.04}$ & $0.13^{+0.03}_{-0.02}$ & $0.13^{+0.01}_{-0.01}$\\[0.1cm]
    $\dot{m}$& $0.2^{+0.016}_{-0.001}$ & $0.011^{+0.001}_{-0.003}$ & $0.1^{+0.01}_{-0.01}$ & $0.11^{+0.02}_{-0.02}$ & $0.14^{+0.02}_{-0.02}$ & $0.1^{+0.01}_{-0.01}$\\[0.1cm]
    $N$  & \multicolumn{6}{|c|}{$0.81^{+0.02}_{-0.04}$(l)}\\[0.1cm]
    \hline    
    $\chi^{2}/{\rm DoF}$ & \multicolumn{6}{|c|}{2017/1573}\\[0.1cm]
    $\chi^{2}$ & 882 & 349 & 216 & 179 & 202 & 189 \\[0.1cm]
    \hline
  \end{tabular}\\
{\it Notes:}
Parameters denoted with '(l)' are linked across S, N, B1, B2, B3 and B4. Parameters denoted with '(f)' are fixed.
For {\tt borus} we fixed the inclination angle of $i = 47 \degr$ \citep{2017MNRAS.472.2170S}. 
 The Galactic absorption is modelled using  \texttt{Tbabs} with $N_{\rm H} = 1.1 \times 10^{20}$ cm$^{-2}$.
 For {\tt kerrflow} $N=1$ corresponds to $M = 8 \times 10^7$ M$_{\odot}$ and  $d = 21.2$ Mpc.
  \label{tab:fit}
\end{table*}

\subsection{NGC 7213}

The X-ray spectrum of NGC 7213 is dominated by intrinsic emission with a weak contribution of other components, therefore, we used the full energy ranges of all detectors.  

We built the spectral model 
with the internal X-ray emission described by {\tt kerrflow} and we included additional components indicated by previous studies of NGC 7213 \citep{2003A&A...407L..21B,2005MNRAS.356..727S,2008MNRAS.389L..52B,2010MNRAS.408..551L}. The definition of our model is given in the header of Table 
\ref{tab:fit}.
We do not find evidence of any internal absorption; the upper limit for the column density of an additional absorber is $\sim 10^{19}$ cm$^{-2}$.
In soft X-rays, NGC 7213 shows signatures of emission by a collisionally ionized thermal plasma with $kT \sim 0.3$ keV, which we describe using a single \texttt{apec} component.
We tried also  a version with two \texttt{apec}  components, for which  we found temperatures, $kT_1=0.27$ keV and $kT_2=0.75$ keV, consistent with these reported in \cite{2010MNRAS.408..551L},
however, this version gives the same fit quality as the one with a single \texttt{apec}. 
We attempted to apply the neutral reprocessing, {\tt borus}, to all 6 datasets, however, adding it for any of the {\it BeppoSAX} spectra does not improve the fit and does not affect the fitted parameters, therefore, for our final model, we use it only for spectrum N and S.   
We use the phenomenological {\tt zgauss} (cosmologically redshifted Gaussian) model to describe the blueshifted Fe XXV and Fe XXVI lines reported in many previous analyses of NGC 7213.
We found these lines to be statistically significant in the spectra N and S and their parameters are consistent with those reported by \cite{2010MNRAS.408..551L} and \cite{2015MNRAS.452.3266U}. Finally, we included an additional line at 8 keV, with parameters reported in \cite{2010MNRAS.408..551L} and we find that it significantly improves the fit but only for spectrum S. We did not find a statistically significant improvement after adding any of these lines to the {\it BeppoSAX} spectra.

For NGC 7213 we normalize the {\tt kerrflow} model to $M = 8 \times 10^7$ M$_{\odot}$ and  $d = 21.2$ Mpc\footnote{The luminosity distance from https://ned.ipac.caltech.edu/} and we allow $N$ to
 fit freely. 
We first linked the values of both $\delta$ and $\beta$ across spectra S, N and B1 -- B4 to check if spectral changes can be reproduced by the change of $\dot m$. This version gives  a rather poor fit with $\chi^2/{\rm DoF} \simeq 1.43$ for the fitted $\beta \simeq 1$, $\delta \simeq 0.01$ and $N \simeq 3.71^{+0.02}_{-0.01}$, which normalization  is in some tension with the estimates of $M$ and $d$.
We then checked if the fit improves when $\delta$ or $\beta$ are allowed to fit freely for each spectrum.
We found a significant improvement, with $\Delta \chi^2 = -204$ for 5 free parameters, in the model with unlinked $\delta$.
Also in favour of this varying--$\delta$ model, the fitted $N \simeq 0.8$ is consistent with the estimated $M$ and $d$. 
We do not find indications for the change of $\beta$, allowing it to fit freely gives $\Delta \chi^2 = -5$ for 5 free parameters.
Then, our model with a changing $\delta$ and a constant $\beta$  gives the best description of the internal X-ray emission of NGC 7213 and we show it in Table \ref{tab:fit} and Figure \ref{fig:spectra}. 
The four spectra measured by {\it BeppoSAX} are very similar to each other (which is not suitable to break the model degeneracies), therefore, restricting our analysis to spectra B1--B4 gives very poorly constrained model parameters.

At the range of parameters fitted in NGC 7213, the model spectra do not exhibit departures from a power-law shape in the X-ray range except for spectrum N, see Figure \ref{fig:sed}b.
Indeed, fitting this spectrum (i.e.\ N) with either a simple power-law or an exponentially cut-off power-law models, we find a significant improvement for the latter, with $\Delta \chi^2 = -41$, for the fitted $E_{\rm cut} = 85^{+31}_{-18}$ keV and $\Gamma = 1.74^{+0.03}_{-0.03}$. In our {\tt kerrflow} model, this softening above $\sim 10$ keV is reproduced by the shape of the second scattering bump. We did not find indications for departures from a power-law of the intrinsic component in the remaining X-ray spectra of NGC 7213. We note that 
this actually sets an upper limit on the value of $\delta$, because the model spectra predicted for $\delta \ga 0.2$ expose deviations from a power-law shape, which are not present in the observed data. For example, fixing $\delta = 0.5$ we obtained a poor fit with $\chi^2/{\rm DoF} = 2351/1579$, with systematic residuals related to the predicted spectral distortions. On the other hand, we can also rule out the model with a constant $\delta = 10^{-3}$. Fixing this value of $\delta$ we get $\chi^2/{\rm DoF} = 2270/1579$, for $N \simeq 3.1$ (i.e.\ in some tension with the estimations of $M$) and $a \simeq 0$; the residuals are most pronounced for spectrum N, which is markedly inconsistent with this low value of $\delta$. On the other hand, allowing $\delta$ to fit freely for all spectra  but fixing $a=0$, we found the fit similar to the one with  $\delta=10^{-3}$ except for spectrum N, for which the fitted $\delta \simeq 0.5$, with $\chi^2/{\rm DoF} = 2163/1574$, i.e.\ still much worse than our best fit with $a \simeq 0.95$.

Similarly as for NGC 4258, we also refitted the model given in Table \ref{tab:fit}, but using an unlinked $a$, to check the rationality of the model. Remarkably, we found consistent values of $a$ in all datasets, namely $a=0.95^{+0}_{-0.06}$ in S, $a=0.95^{+0}_{-0.06}$ in N, $a=0.95^{+0}_{-0.09}$ in B1, $a=0.95^{+0}_{-0.09}$ in B2, $a=0.95^{+0}_{-0.33}$ in B3 and $a=0.95^{+0}_{-0.24}$ in B4, with strong indications of a rapid rotation of the supermassive black hole in all cases.

\subsection{$\gamma$-ray emission}

In Figure \ref{fig:sed} we show the results of our analysis of the LAT data. At the location of NGC 4258, we found a marginal signal in the 0.1--1 keV range, with the test statistic $= 30$ (corresponding to $\sigma \sim 5$). We found for it the photon spectral index $\Gamma = 4.16 \pm 0.49$ and the photon flux $F_{\gamma} = (6.32 \pm 1.28) \times 10^{-9}$ phot cm$^{-2}$ s$^{-1}$. However, it is not possible to assess if this signal represents a real emission from NGC 4258, due to the poor angular resolution of LAT at these energies. At $E>1$ GeV, we did not find a statistically significant signal. Also in NGC 7213 we did not find any significant signal in the whole LAT energy range. Then, we derived the upper limits for the photon flux at the 95 per cent confidence level, assuming photon indices appropriate for the $\pi^0$-decay photon spectra. For the 0.2--1 GeV range in NGC 7213 we found $F_{\gamma} < 1.7 \times 10^{-9}$ phot cm$^{-2}$ s$^{-1}$ for the assumed $\Gamma=4$. For the 1--10 GeV range in both NGC 4258 and NGC 7213 we found $F_{\gamma} < 3 \times 10^{-11}$ phot cm$^{-2}$ s$^{-1}$, insignificantly dependent on $\Gamma$ in the range  $2.1 \le \Gamma \le 2.7$.
%($\Gamma = 2.1$: $F_{\gamma} < 2.2$; $\Gamma = 2.7$: $F_{\gamma} < 3.3$).($\Gamma = 2.1$: $F_{\gamma} < 2.9$; $\Gamma = 2.7$: $F_{\gamma} < 3.3$) 

We use average parameters of the flow indicated by our modelling of the X-ray spectra, i.e.\ $\dot m = 0.003$, $\delta=0.1$, $\beta=1$ and $a=0.6$ for NGC 4258, and  
$\dot m = 0.1$, $\delta=0.1$, $\beta=3$ and $a=0.95$ for NGC 7213, and we apply the model developed by \citet{2013MNRAS.432.1576N} and \citet{2015A&A...584A..20W} to compute the predicted $\gamma$-ray fluxes. We use the internal energy of protons from our GR hydrodynamic solutions for the above parameters. We compute the rest-frame spectra of $\gamma$-rays emitted through the production and decay of neutral pions, assuming that the proton energy distribution has a high energy tail in the form of a power law with an index $s$, containing a fraction $\eta$ of the total energy of protons, and that the remaining part, $1 - \eta$, is stored in  protons whose energy distribution is thermal. We then compute the GR transfer of $\gamma$-rays, taking into account their absorption in $\gamma \gamma$ interactions with lower energy photons of the radiation field tabulated in the simulation of Comptonization within the flow. The example spectra for $s=2.4$ are shown in Figure \ref{fig:sed}. 

For NGC 4258 the predicted $\gamma$-ray flux is much below the LAT sensitivity for any parameters of the proton distribution. For NGC 7213 the comparison with the LAT upper limits allows us to constrain the content of nonthermal protons to $\eta \la 0.1$. We note also that the internal $\gamma \gamma$ absorption strongly attenuates the $\gamma$-ray flux observed from hot flows; neglecting it for NGC 7213 we get the signal above the LAT upper limits, see Figure \ref{fig:sed}b.

\begin{figure*}
\centering
\includegraphics[height=5.7cm]{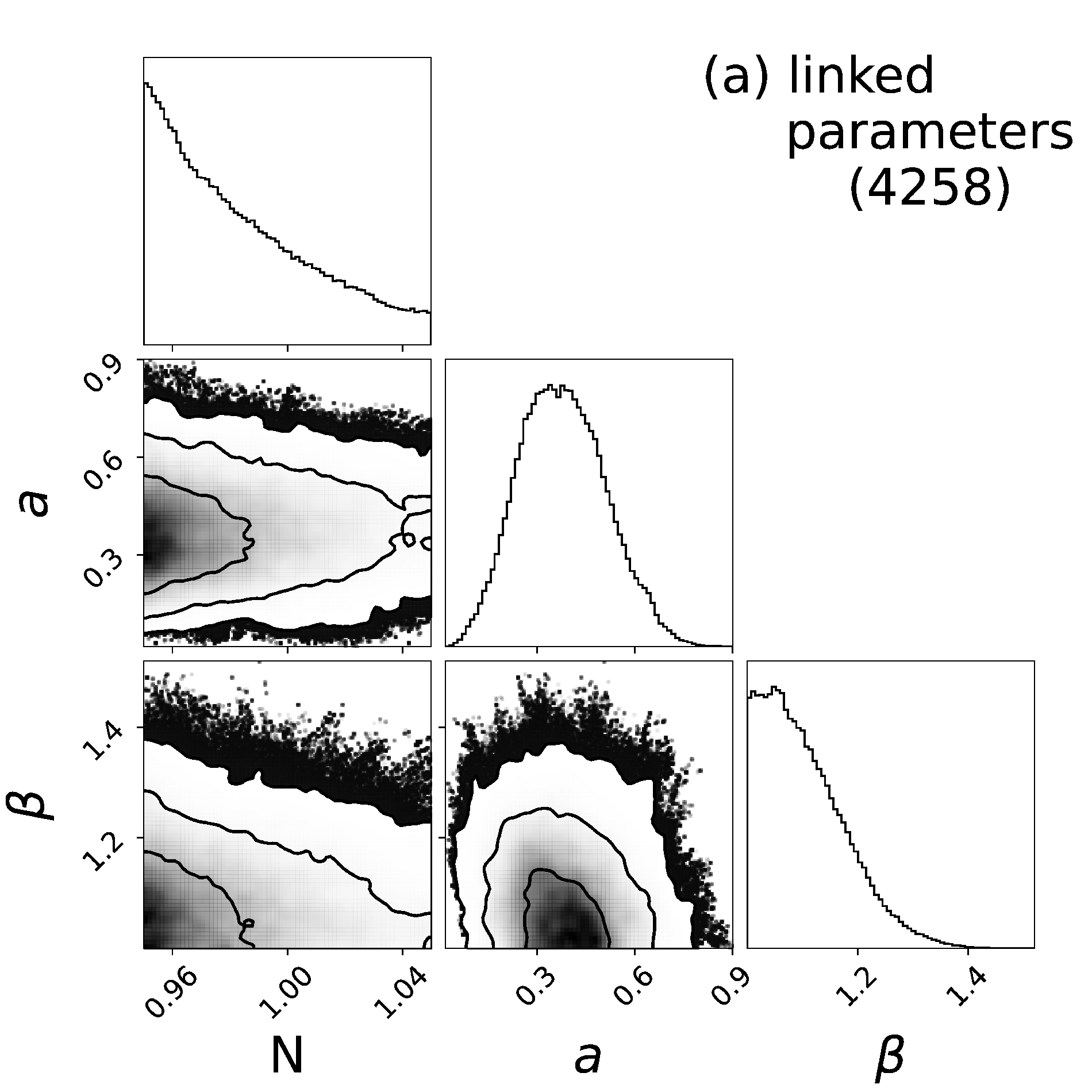}
\includegraphics[height=5.7cm]{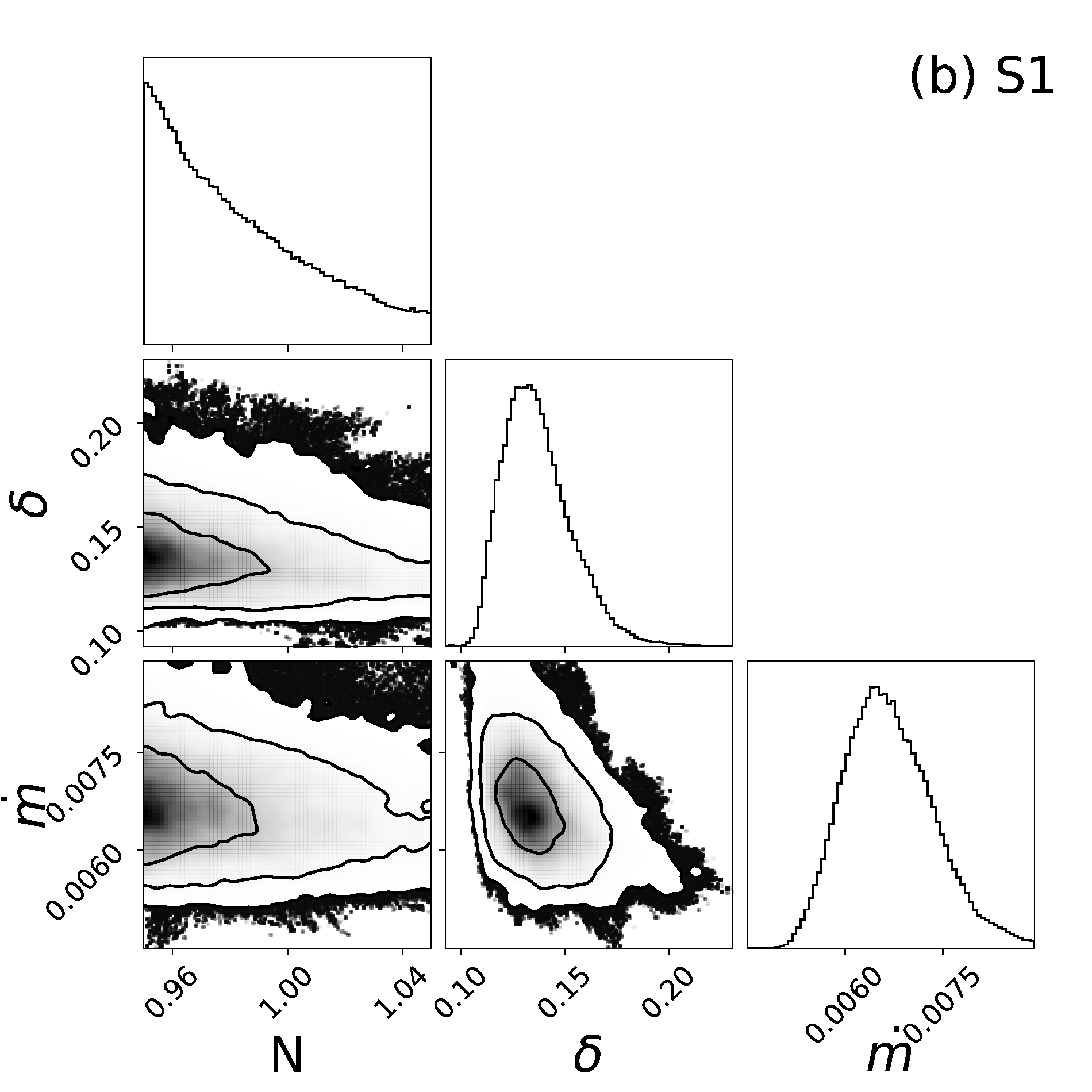}
\includegraphics[height=5.7cm]{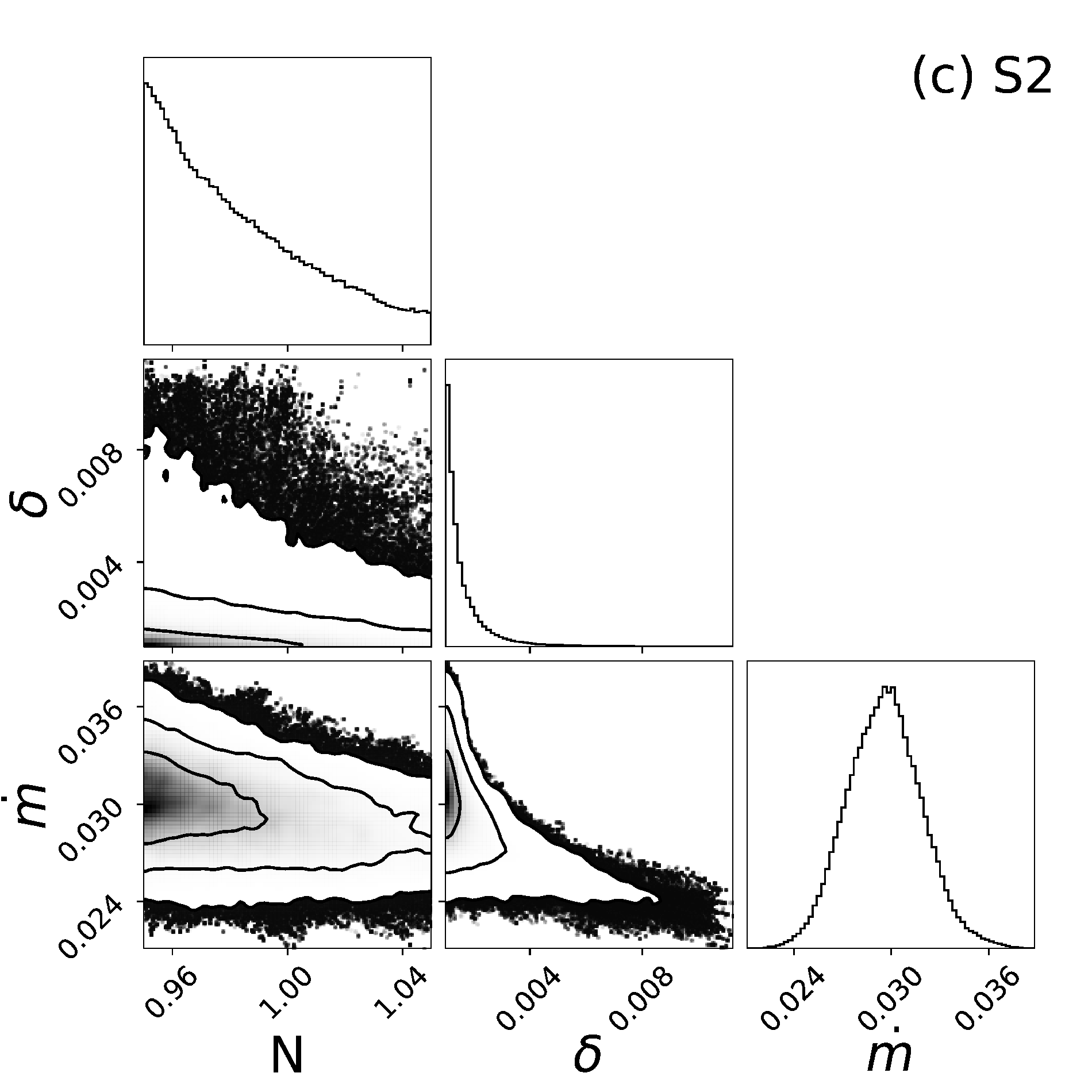}\\
\includegraphics[height=5.7cm]{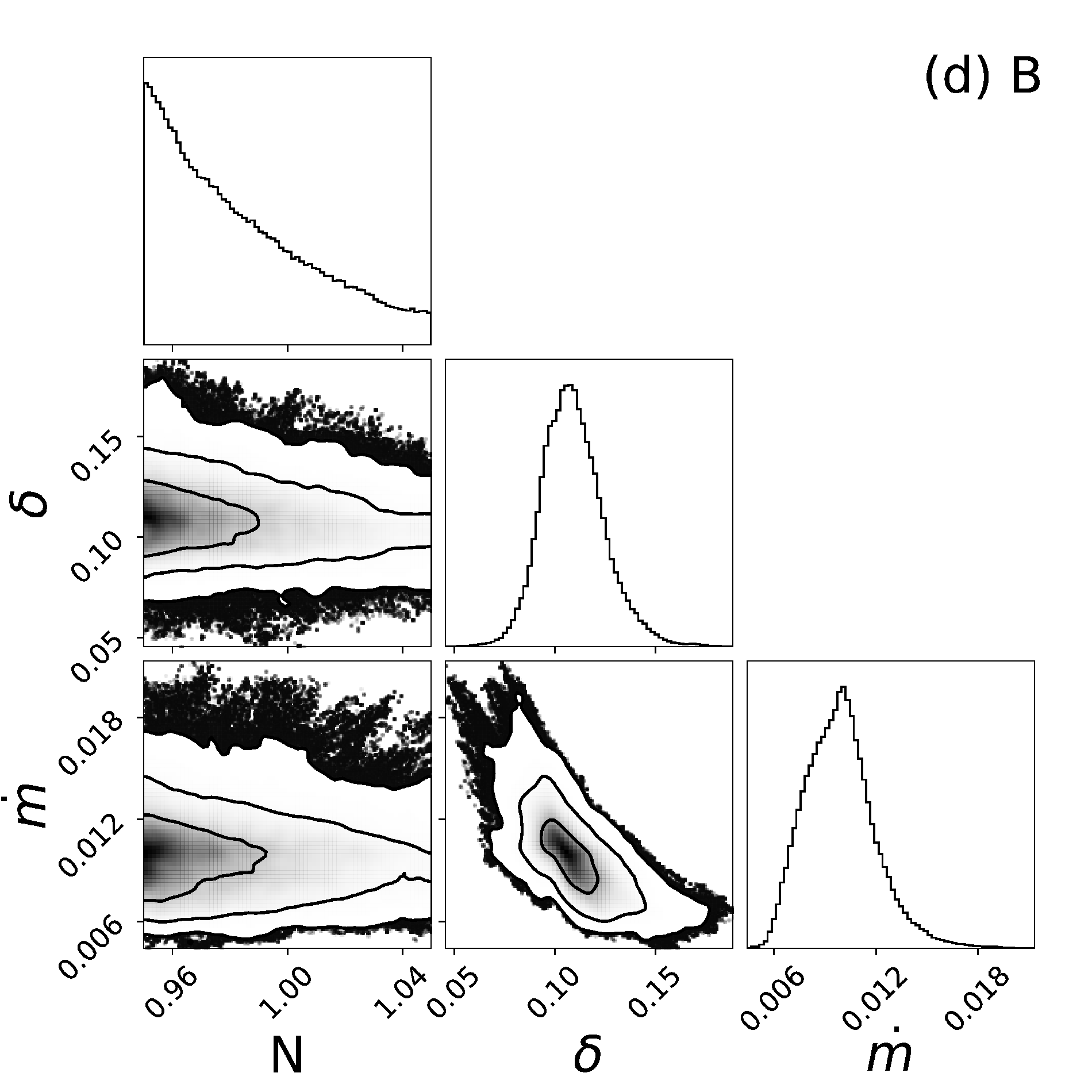}
\includegraphics[height=5.7cm]{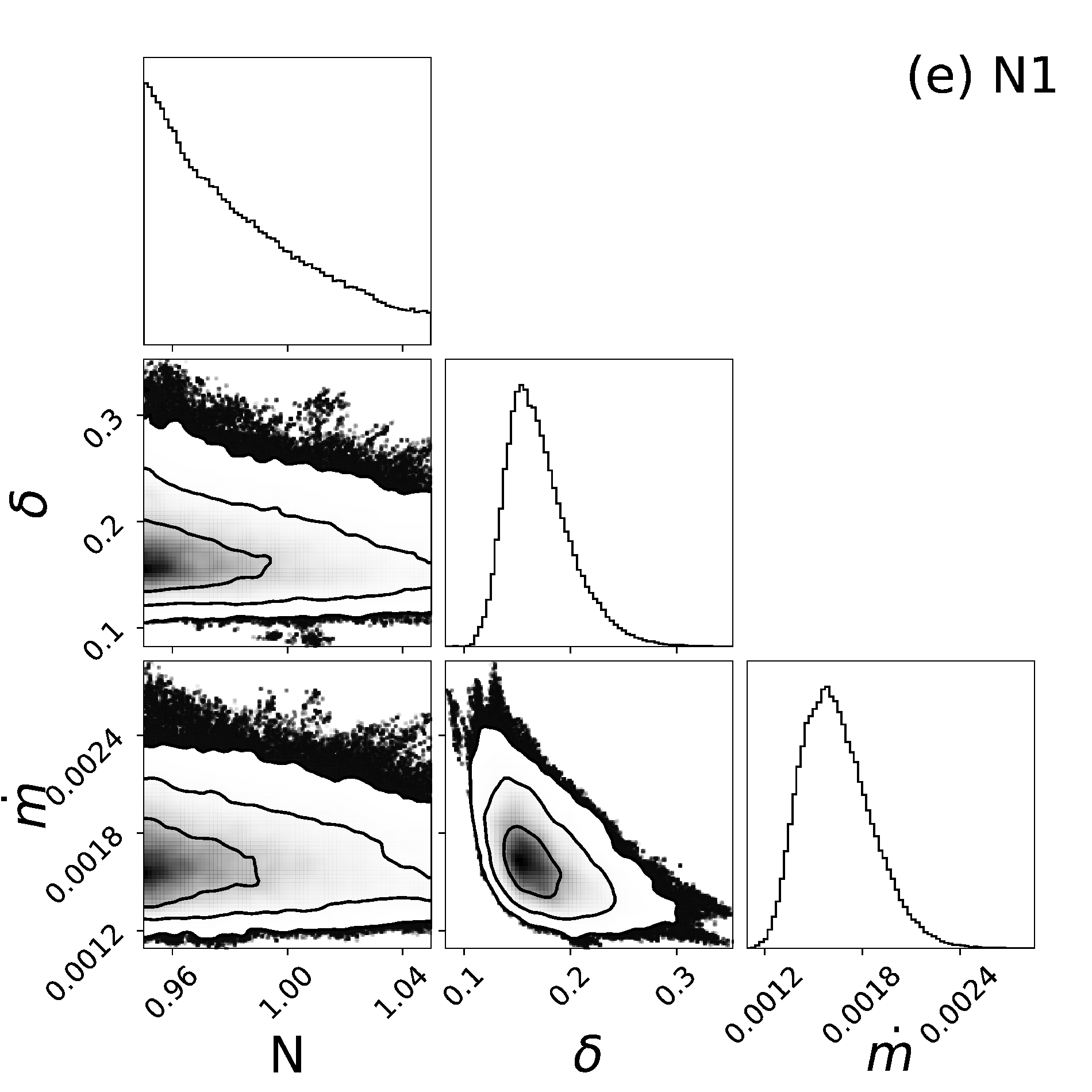}
\includegraphics[height=5.7cm]{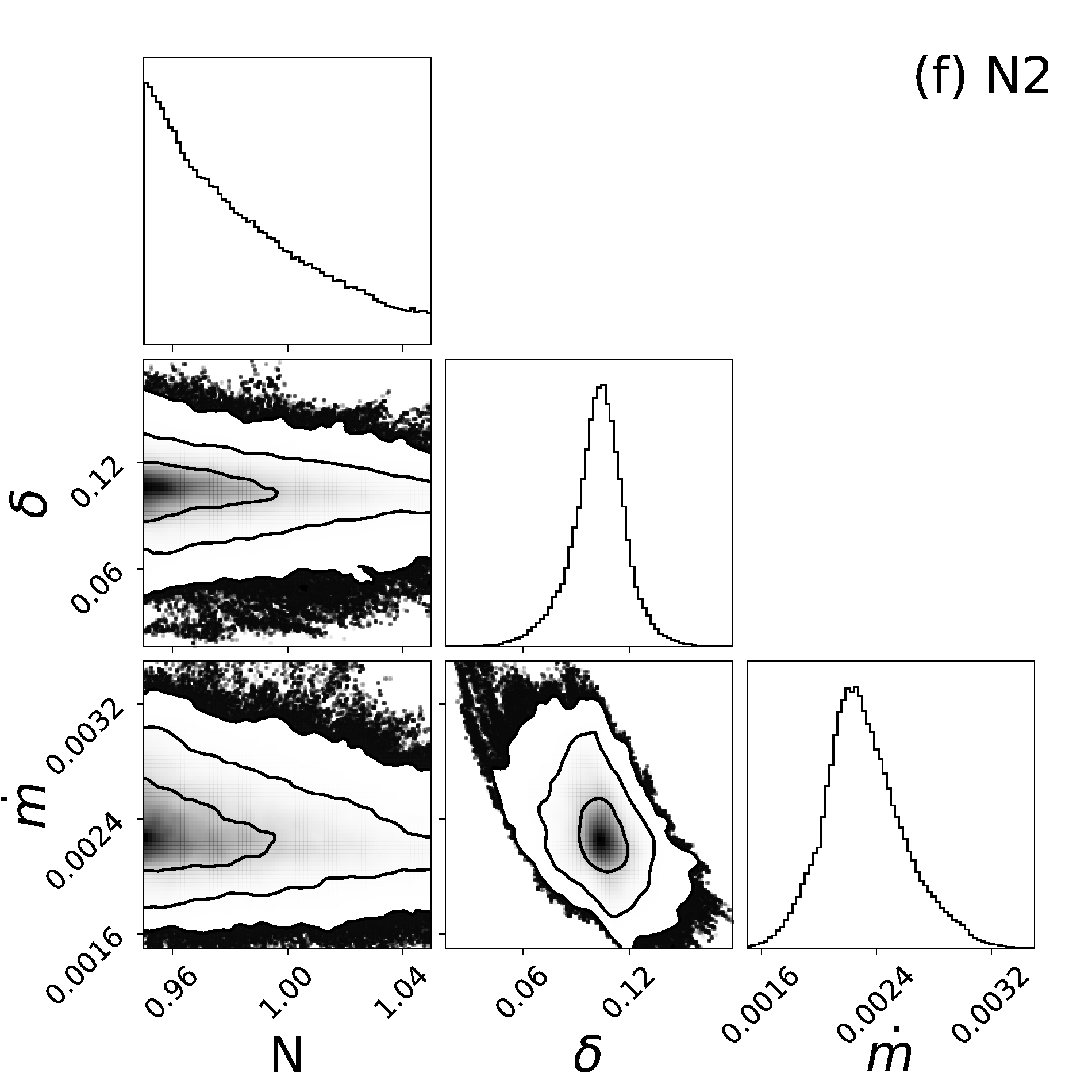}\\
\includegraphics[height=5.7cm]{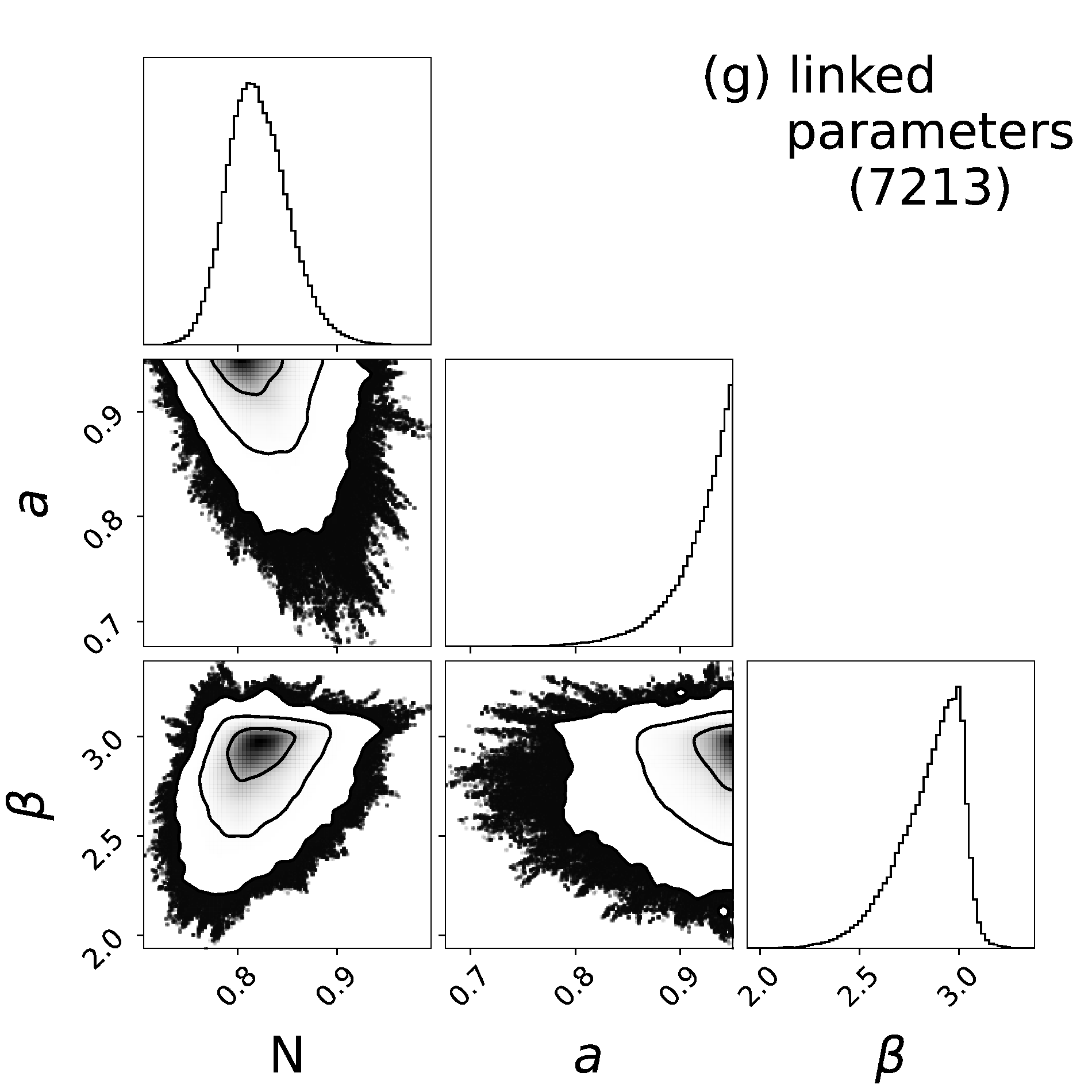}
\includegraphics[height=5.7cm]{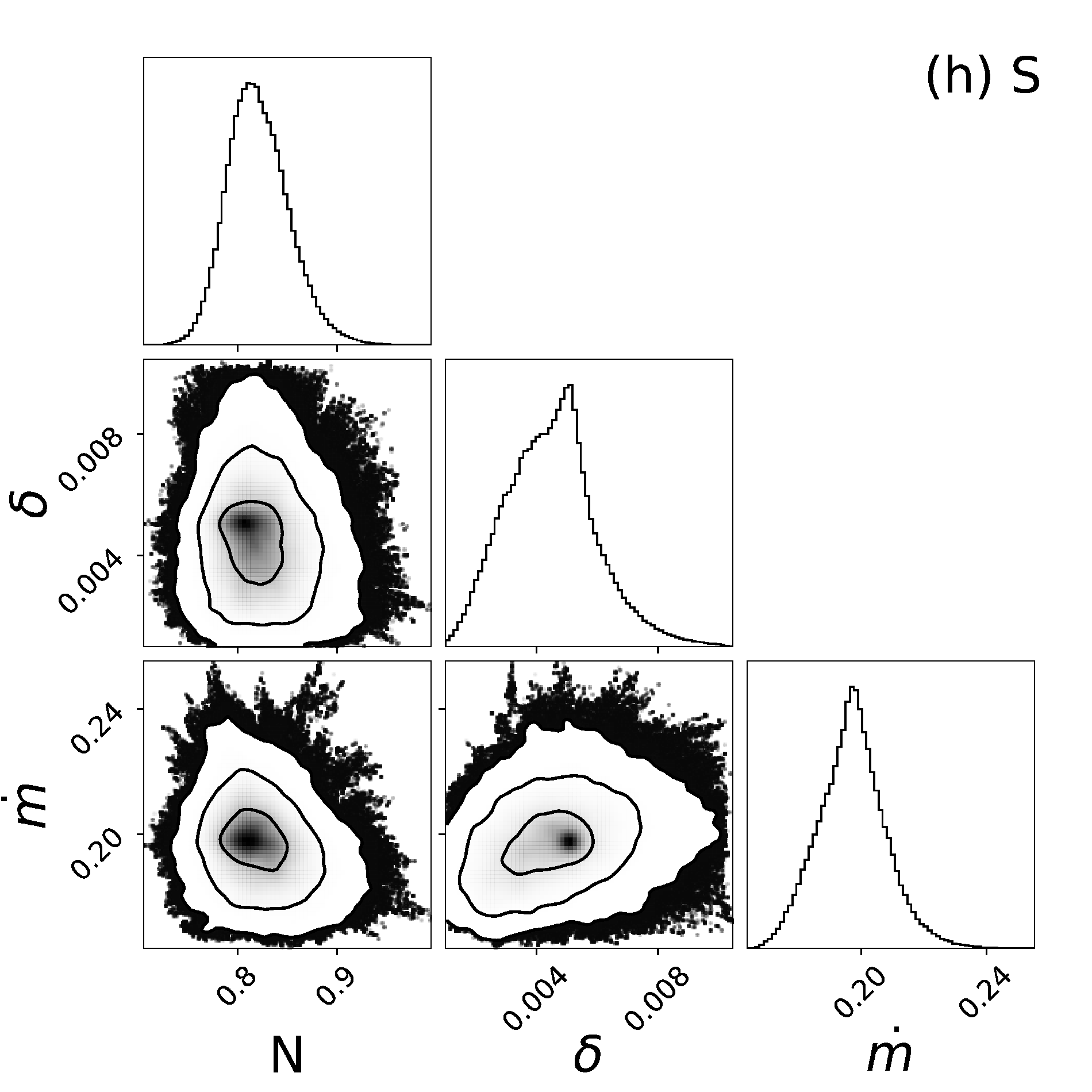}
\includegraphics[height=5.7cm]{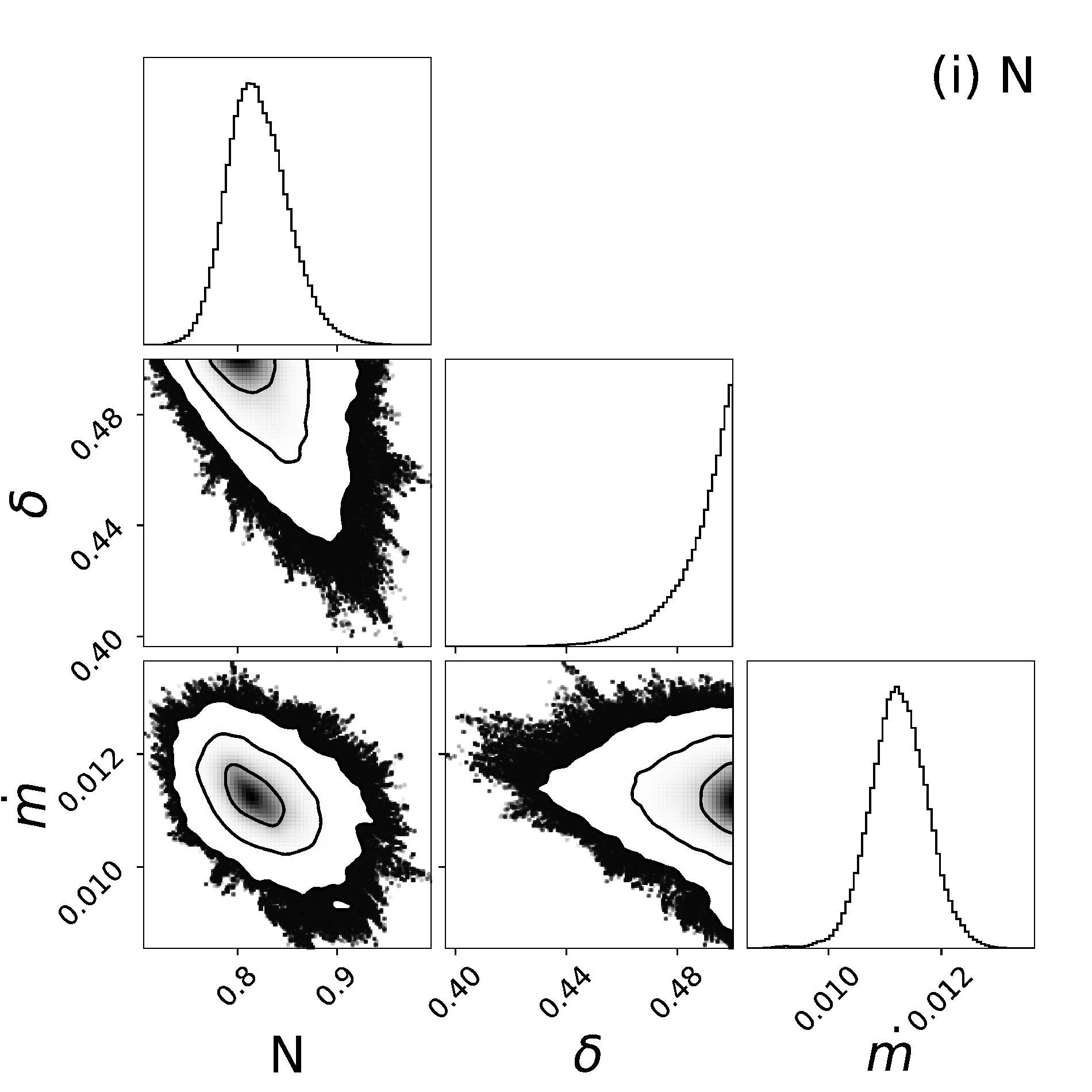}\\
\includegraphics[height=5.7cm]{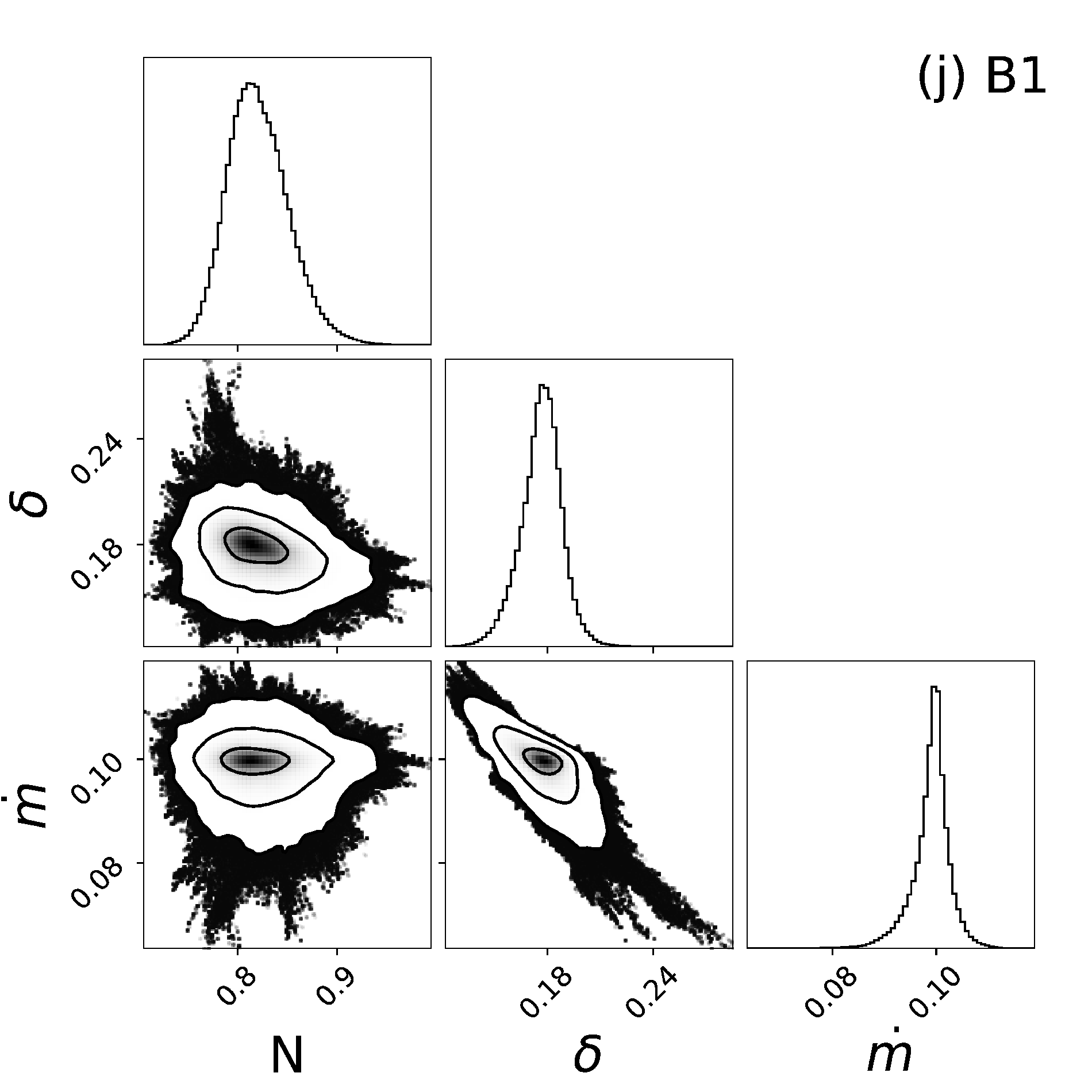}
\includegraphics[height=5.7cm]{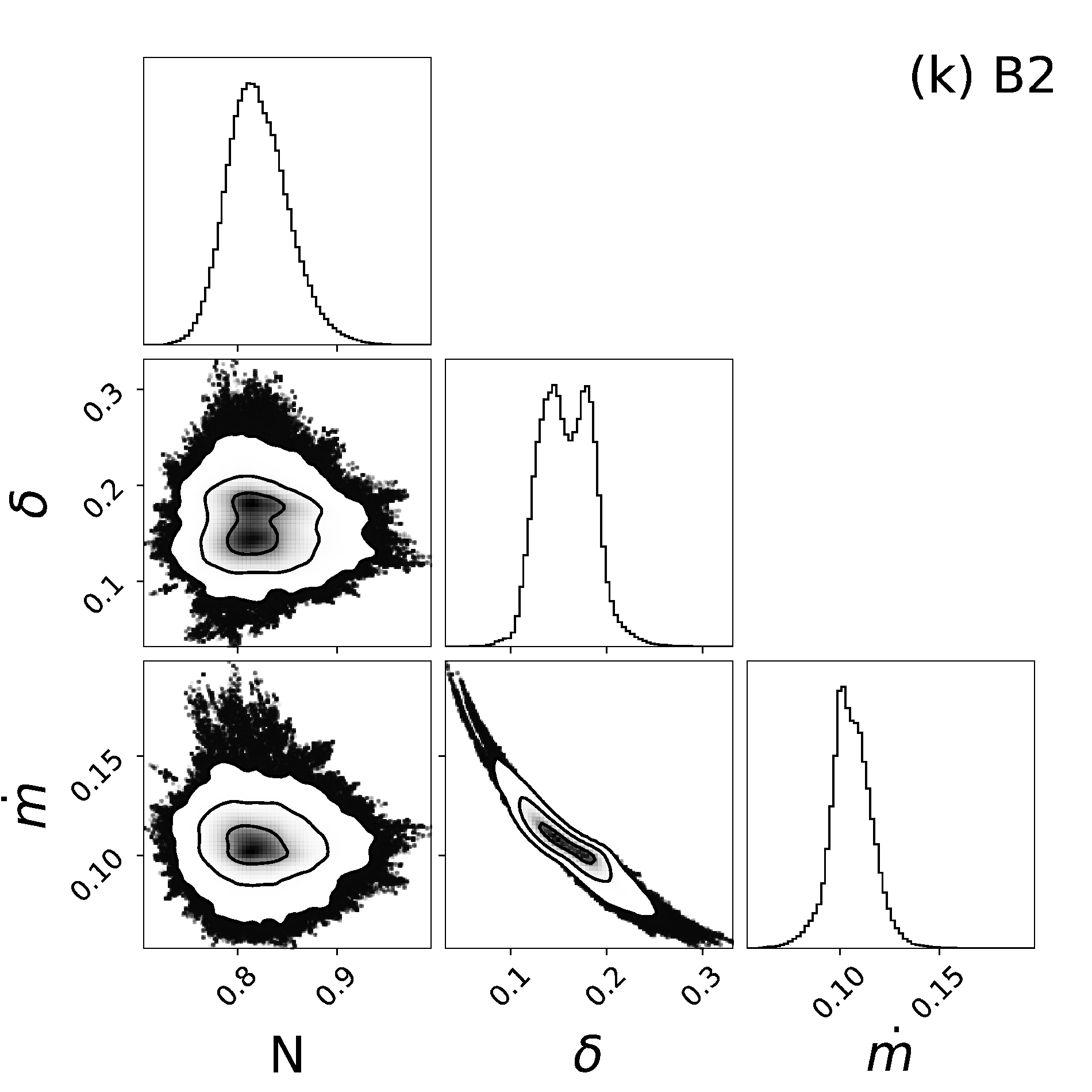}
\includegraphics[height=5.7cm]{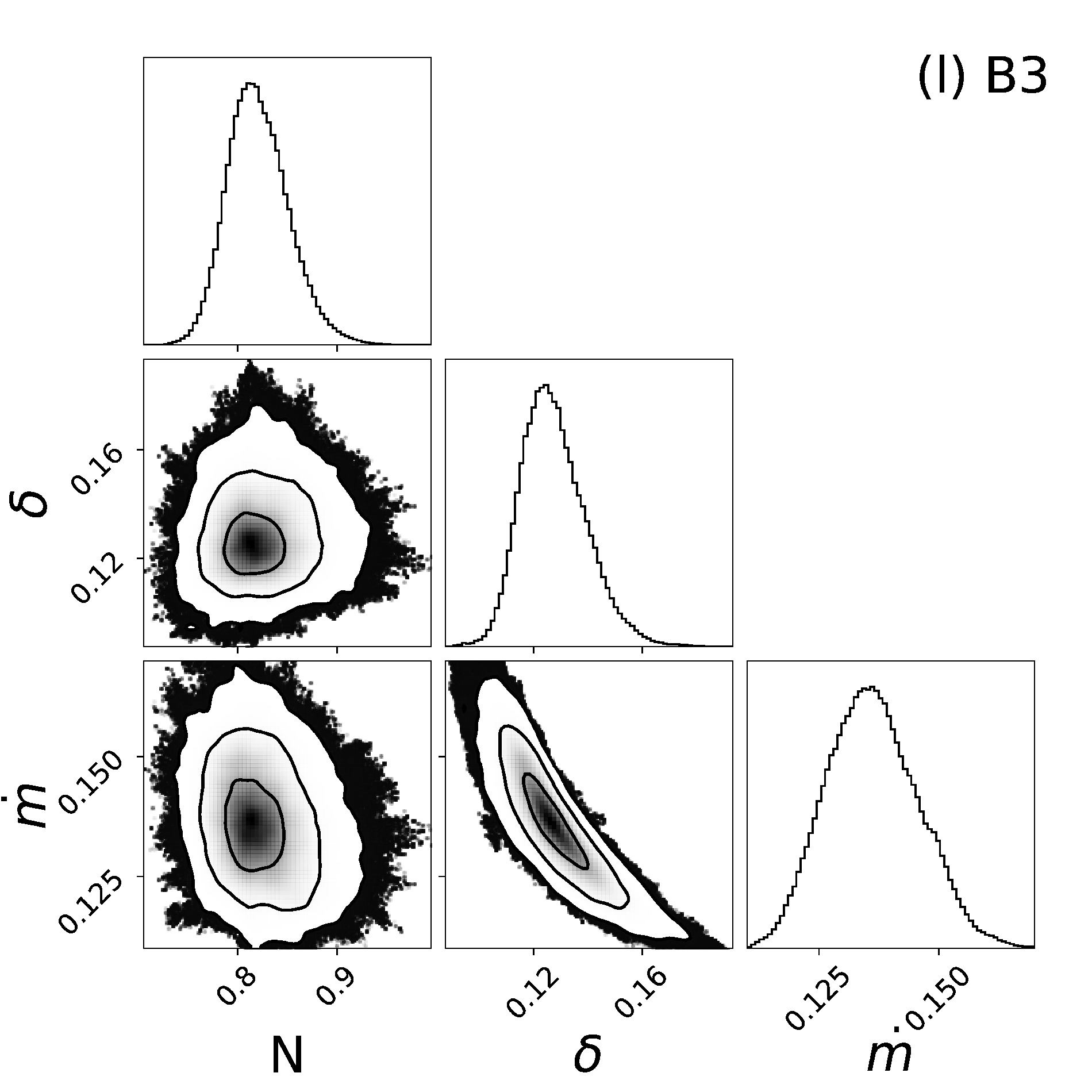}
\caption{
Probability distributions showing correlations between parameters fitted to (a-f) NGC 4258, see Table \ref{tab:fit_4258}, model with a free $\delta$, and (g-l) NGC 7213,
see Table \ref{tab:fit}, obtained in the MCMC analysis using \texttt{xspec\_emcee} implemented by Jeremy Sanders. The results of this analysis are presented using package \texttt{corner}. The contours in the 2D plots correspond to the significance of $\sigma = 1, 2, 3$. The histograms show the probability distributions for the individual parameters. (a) and (g) show parameters $N$, $a$ and $\beta$, linked across all fitted spectra; (b-f) and (h-l) show parameters $\dot{m}$ and $\delta$ fitted to the individual spectra, as labelled on the panels, and $N$.   
}
\label{fig:mcmc}
\end{figure*}

\begin{figure}
\centering
 \includegraphics[height=5cm]{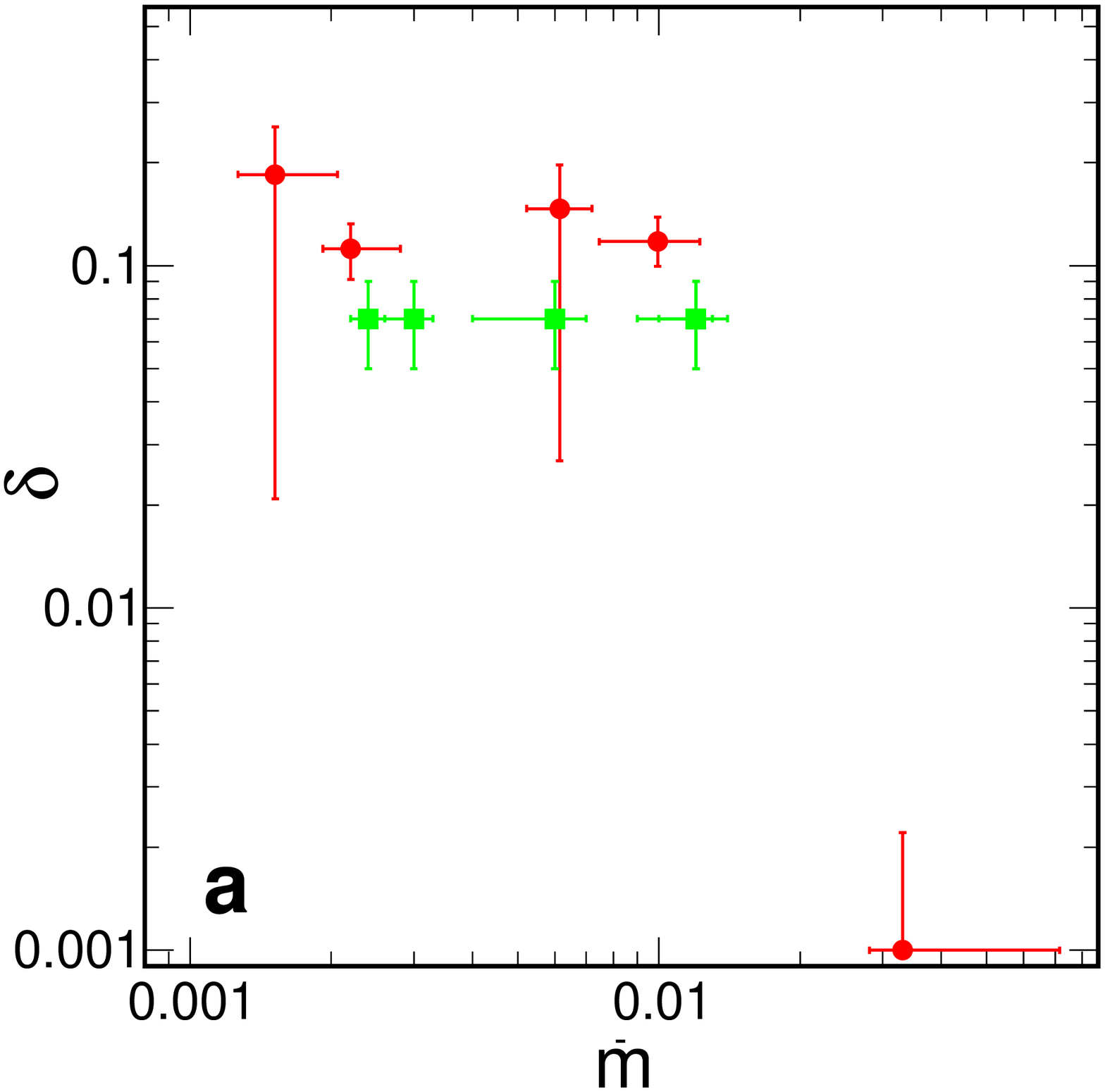}
 \includegraphics[height=5cm]{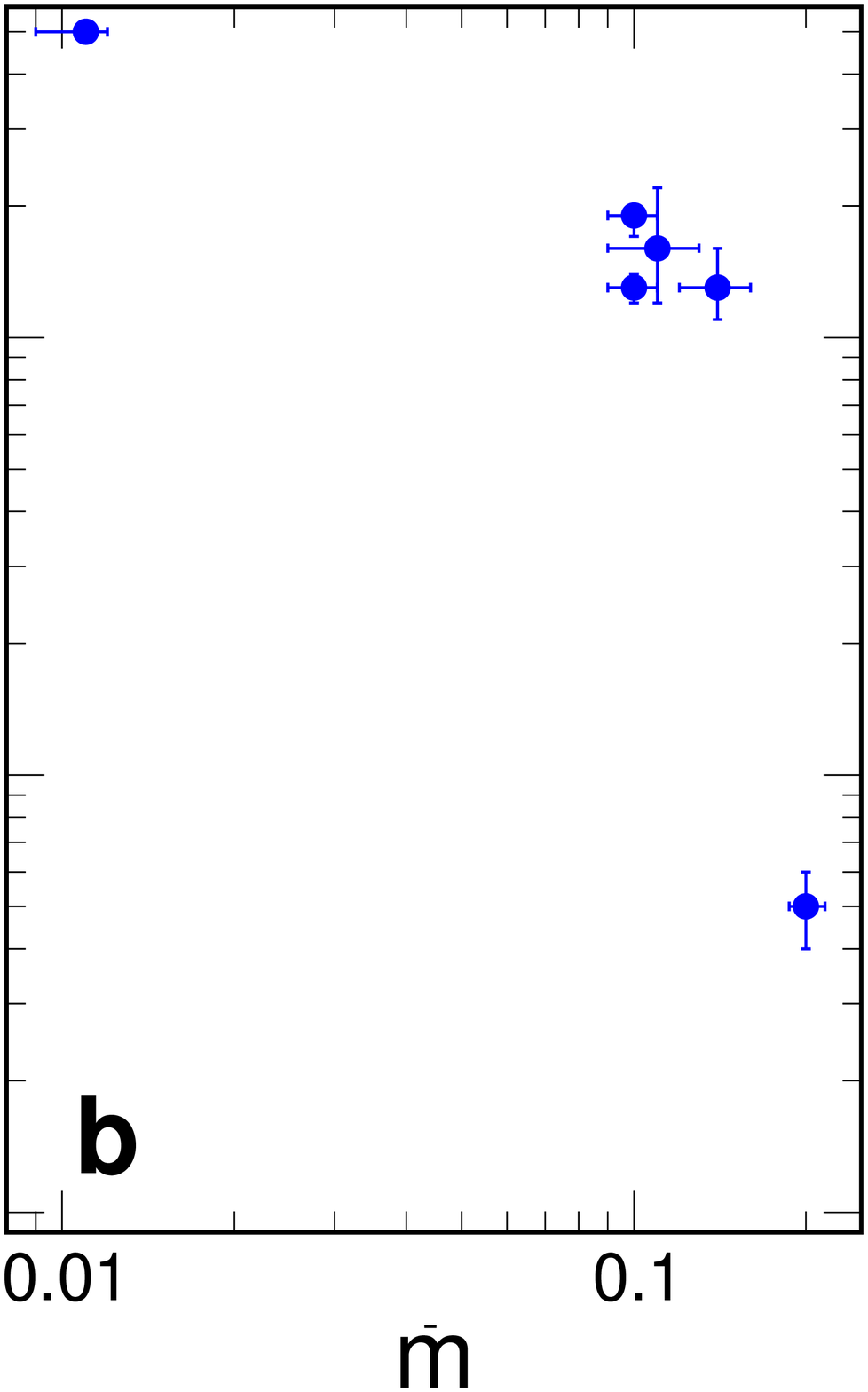}
\caption{The fitting results in the $\dot m - \delta$ plane for a) NGC 4258 and b) NGC 7213, as given in Tables \ref{tab:fit_4258} and \ref{tab:fit}, respectively.
For NGC 4258 we show results for both the version with a linked (green squares) and unlinked (red circles) $\delta$.}
\label{fig:mdot_delta}
\end{figure}

\section{Summary and discussion}
\label{sect:discuss}

Our main results are summarized as follows:

\noindent
(i) We applied a self-consistent, GR model to estimate the parameters of hot flows in two well-studied AGNs, NGC 4258 and NGC 7213. We do not find any significant degeneracies in the model (except for a minor anticorrelation between $\dot m$ and $\delta$ in some spectra, see Figure \ref{fig:mcmc}, which does not lead to significant uncertainty in these parameter values),
and in both objects, both $\delta$ and $\beta$ are tightly constrained. These two parameters are of major importance for modelling the hot-flow spectra, as they determine the electron heating and cooling efficiencies, and yet they are poorly constrained by either theory or observations. Two notable previous attempts to estimate $\delta$ include  \citet{2003ApJ...598..301Y}, who favour $\delta \simeq 0.5$ from modelling of the quiescent emission of Sgr A*, and the study of 12 AGNs by \citet{2014MNRAS.438.2804N} in which a large of range of $\delta$ between 0.01 and 0.3 was found. Both these works, however, used a non-GR hot-flow model with an incomplete \citep[i.e.\ local and non-relativistic approximation of][]{1990MNRAS.245..453C} treatment of thermal Comptonization.
The influence of the spin parameter is comparatively weaker, nevertheless, it determines the flow dynamics which then affects the details of spectral changes induced by the change of model parameters. We found a significant hint for a rapid rotation, with $a \ga 0.9$, of the supermassive black hole in NGC 7213. In NGC 4258 we find some preference for a moderate black-hole spin with $a \sim 0.5$.

\noindent
(ii) In both objects, we found similar values of plasma magnetization, close to equipartition with the gas pressure ($\beta \simeq 1$ in NGC 4258 and $\beta \simeq 3$ in NGC 7213), which appears constant for all analysed spectra, and similar electron heating efficiencies, $\delta \simeq 0.1$, indicated by most data sets, which, however, shows significant variations in some cases. Especially in NGC 7213 a model with a constant $\delta$ is clearly inconsistent with the studied X-ray data sets.
These changes of $\delta$ appear to be negatively correlated with $\dot m$, see Figure \ref{fig:mdot_delta}. 
We presume that this apparent behaviour may be due to local density fluctuations, which dominate the instantaneous production of radiation, whereas the heating rate still corresponds to the average accretion rate. Such a fluctuation in our model would be represented by a change of $\dot m$ (which determines the optical depth) and, if the actual heating rate remains unchanged, by an apparent change of $\delta$.

\noindent
(iii) The electron temperature in hot flows is relativistic and their Comptonization spectra extend to the MeV range, see Figure \ref{fig:sed}. This prediction currently cannot be verified observationally; future soft $\gamma$-ray detectors, like COSI or e-ASTROGAM, would be essential for this purpose. The currently available data, at $\la 100$ keV, for NGC 4258 and NGC 7213 are fully consistent with predictions of the hot-flow model for relevant accretion rates. 
For NGC 4258, 
\citet{1999ApJ...516..177G} estimated the accretion rate of $\dot M \ga 1.5 \times 10^{-4} M_{\odot}$ yr$^{-1}$ by modelling the disc at $>0.1$ pc, at which the viscous time-scale is of the order of a gigayear. The accretion rate in the inner region may be different as the mass flux may change significantly over this timescale, nevertheless, the accretion rate found in our model, $\dot M = (1.8 - 8) \times 10^{-4} M_{\odot}$ yr$^{-1}$ (for $\dot m = 0.002 - 0.01$), is in a remarkable agreement with that estimate. 
As noted by \citet{2022A&A...663A..87M}, the intrinsic luminosity of NGC 4258 displayed a long-term decrease between the early 2000s and 2015, which in our modelling is explained by the decrease of the accretion rate by a factor of $\sim 4$. 

\noindent 
(iv) Changes in $\dot m$, driving the spectral changes in our models, may be due to the viscous evolution of the inner flow, occurring at the viscous timescale, $t_{\rm visc} = [\Omega_{\rm K} \alpha (H/R)^2]^{-1}$, where $\Omega_{\rm K}$ is the Keplerian angular velocity, $\alpha$ is the viscosity parameter and the aspect ratio, $H/R$, is $\simeq 1$ in our hydrodynamical solutions at $R \sim (100 - 10000) R_{\rm g}$, where $R_{\rm g} = GM/c^2$ is the gravitational radius. In all fitted models, the bulk of the X-ray radiation is produced within the central $100 R_{\rm g}$. In both objects we estimated large changes of $\dot m$ on a timescale of 5 years, which for both sources corresponds to the viscous timescale at $R > 2000 R_{\rm g}$, much longer than $t_{\rm visc}$ in the emitting region.  
On the other hand, only small variations of $\dot m$, by a factor of $\la 1.5$, are found in NGC 4258 on a two-month timescale (spanned by observations N1 and N2), corresponding to $t_{\rm visc}$ at $\simeq 400 R_{\rm g}$ and in NGC 7213 on a two-year timescale (spanned by observations B1--B4), corresponding to $t_{\rm visc}$ at $\simeq 1300 R_{\rm g}$.

\noindent
(v) In our solutions with $\delta \simeq 0.1$ for both objects, the radiative efficiency, defined as $\eta = L/(\dot M c^2)$, where $L$ is the bolometric luminosity, is $\eta \simeq 0.02$.

\noindent
(vi)
In NGC 7213, the comparison of the model prediction with the LAT data allows us to constrain the content of nonthermal protons to $\la 10$ per cent. A similar constraint was found in \citet{2015A&A...584A..20W}, however, that work assumed $\delta = 10^{-3}$ (which maximizes the $\gamma$-ray signal as it corresponds to the largest density of the flow), whereas here we used the much larger value of $\delta$ measured from the X-ray data. We also used here updated LAT upper limits, by a factor of several lower than those found by \citet{2015A&A...584A..20W}.

\noindent
(vii)
In agreement with previous findings \citep[e.g.][]{2003A&A...407L..21B}, in both AGNs, the Fe K$\alpha$ line is found to be emitted by a Compton-thin material, with column density $\simeq 10^{23}$ cm$^{-2}$, possibly in the broad line region \citep{2008MNRAS.389L..52B}. 
The column density of the internal absorber in NGC 4258 is of the same order, $\simeq 10^{23}$ cm$^{-2}$, and \citet{2022A&A...663A..87M} suggested periodic fluctuations of this obscuring column density on a timescale of 10 years. We note that the {\it Suzaku} observation in 2010, which was neglected in their work, disagrees with such a periodic change, according to which the column density in 2010 should be low, $\sim 6 \times 10^{22}$ cm$^{-2}$, whereas the measured value is larger by over a factor of 2.

\section*{ACKNOWLEDGEMENTS}
We thank Chris Reynolds, the referee, for valuable comments and Mauro Orlandini for help with the {\it BeppoSAX} data. This research has been supported in part by the Polish National Science Centre grants 2015/18/A/ST9/00746, 2016/21/B/ST9/02388 and 2019/35/B/ST9/03944.

\section*{Data Availability}
The data are publicly available at NASA's HEASARC. {\tt kerrflow} is available at \url{https://wfis.uni.lodz.pl/kerrflow}

\label{lastpage}


\begin{thebibliography}{}


\bibitem[\protect\citeauthoryear{Abdollahi et al.}{2020}]{2020ApJS..247...33A} Abdollahi S., Acero F., Ackermann M., Ajello M., Atwood W.~B., Axelsson M., Baldini L., et al., 2020, ApJS, 247, 33. doi:10.3847/1538-4365/ab6bcb


\bibitem[\protect\citeauthoryear{Balokovi{\'c} et al.}{2018}]{2018ApJ...854...42B} Balokovi{\'c} M., Brightman M., Harrison F.~A., Comastri A., Ricci C., Buchner J., Gandhi P., et al., 2018, ApJ, 854, 42. doi:10.3847/1538-4357/aaa7eb

\bibitem[\protect\citeauthoryear{Bauer et al.}{2015}]{2015ApJ...812..116B} Bauer F.~E., Ar{\'e}valo P., Walton D.~J., Koss M.~J., Puccetti S., Gandhi P., Stern D., et al., 2015, ApJ, 812, 116. doi:10.1088/0004-637X/812/2/116


\bibitem[\protect\citeauthoryear{Bianchi et al.}{2003}]{2003A&A...407L..21B} Bianchi S., Matt G., Balestra I., Perola G.~C., 2003, A\&A, 407, L21. doi:10.1051/0004-6361:20031054

\bibitem[\protect\citeauthoryear{Bianchi et al.}{2008}]{2008MNRAS.389L..52B} Bianchi S., La Franca F., Matt G., Guainazzi M., Jimenez Bail{\'o}n E., Longinotti A.~L., Nicastro F., et al., 2008, MNRAS, 389, L52. doi:10.1111/j.1745-3933.2008.00521.x

\bibitem[\protect\citeauthoryear{Coppi \& Blandford}{1990}]{1990MNRAS.245..453C} Coppi P.~S., Blandford R.~D., 1990, MNRAS, 245, 453


\bibitem[\protect\citeauthoryear{Emmanoulopoulos et al.}{2012}]{2012MNRAS.424.1327E} Emmanoulopoulos D., Papadakis I.~E., McHardy I.~M., Ar{\'e}valo P., Calvelo D.~E., Uttley P., 2012, MNRAS, 424, 1327. doi:10.1111/j.1365-2966.2012.21316.x

\bibitem[\protect\citeauthoryear{Fiore et al.}{2001}]{2001ApJ...556..150F} Fiore F., Pellegrini S., Matt G., Antonelli L.~A., Comastri A., della Ceca R., Giallongo E., et al., 2001, ApJ, 556, 150. doi:10.1086/321530


\bibitem[\protect\citeauthoryear{Gammie, Narayan, \& Blandford}{1999}]{1999ApJ...516..177G} Gammie C.~F., Narayan R., Blandford R., 1999, ApJ, 516, 177. doi:10.1086/307089

\bibitem[\protect\citeauthoryear{Ho}{2008}]{2008ARA&A..46..475H} Ho L.~C., 2008, ARA\&A, 46, 475. doi:10.1146/annurev.astro.45.051806.110546

\bibitem[\protect\citeauthoryear{Humphreys et al.}{2013}]{2013ApJ...775...13H} Humphreys E.~M.~L., Reid M.~J., Moran J.~M., Greenhill L.~J., Argon A.~L., 2013, ApJ, 775, 13. doi:10.1088/0004-637X/775/1/13


\bibitem[\protect\citeauthoryear{Ichimaru}{1977}]{1977ApJ...214..840I} Ichimaru S., 1977, ApJ, 214, 840. doi:10.1086/155314

\bibitem[\protect\citeauthoryear{Kawamuro et al.}{2016}]{2016ApJ...831...37K} Kawamuro T., Ueda Y., Tazaki F., Terashima Y., Mushotzky R., 2016, ApJ, 831, 37. doi:10.3847/0004-637X/831/1/37

\bibitem[\protect\citeauthoryear{Laha et al.}{2020}]{2020ApJ...897...66L} Laha S., Markowitz A.~G., Krumpe M., Nikutta R., Rothschild R., Saha T., 2020, ApJ, 897, 66. doi:10.3847/1538-4357/ab92ab


\bibitem[\protect\citeauthoryear{Lasota et al.}{1996}]{1996ApJ...462..142L} Lasota J.-P., Abramowicz M.~A., Chen X., Krolik J., Narayan R., Yi I., 1996, ApJ, 462, 142. doi:10.1086/177137

\bibitem[\protect\citeauthoryear{Li et al.}{2009}]{2009ApJ...699..513L} Li Y.-R., Yuan Y.-F., Wang J.-M., Wang J.-C., Zhang S., 2009, ApJ, 699, 513. doi:10.1088/0004-637X/699/1/513


\bibitem[\protect\citeauthoryear{Lobban et al.}{2010}]{2010MNRAS.408..551L} Lobban A.~P., Reeves J.~N., Porquet D., Braito V., Markowitz A., Miller L., Turner T.~J., 2010, MNRAS, 408, 551. doi:10.1111/j.1365-2966.2010.17143.x

\bibitem[\protect\citeauthoryear{Madsen et al.}{2017}]{2017AJ....153....2M} Madsen K.~K., Beardmore A.~P., Forster K., Guainazzi M., Marshall H.~L., Miller E.~D., Page K.~L., et al., 2017, AJ, 153, 2. doi:10.3847/1538-3881/153/1/2


\bibitem[\protect\citeauthoryear{Magdziarz \& Zdziarski}{1995}]{1995MNRAS.273..837M} Magdziarz P., Zdziarski A.~A., 1995, MNRAS, 273, 837. doi:10.1093/mnras/273.3.837

\bibitem[\protect\citeauthoryear{Mahadevan, Narayan, \& Krolik}{1997}]{1997ApJ...486..268M} Mahadevan R., Narayan R., Krolik J., 1997, ApJ, 486, 268. doi:10.1086/304499

\bibitem[\protect\citeauthoryear{Mahadevan}{1999}]{1999MNRAS.304..501M} Mahadevan R., 1999, MNRAS, 304, 501. doi:10.1046/j.1365-8711.1999.02355.x

\bibitem[\protect\citeauthoryear{Manmoto, Mineshige, \& Kusunose}{1997}]{1997ApJ...489..791M} Manmoto T., Mineshige S., Kusunose M., 1997, ApJ, 489, 791. doi:10.1086/304817


\bibitem[\protect\citeauthoryear{Masini et al.}{2022}]{2022A&A...663A..87M} Masini A., Wijesekera J.~V., Celotti A., Boorman P.~G., 2022, A\&A, 663, A87. doi:10.1051/0004-6361/202243231


\bibitem[\protect\citeauthoryear{Miyoshi et al.}{1995}]{1995Natur.373..127M} Miyoshi M., Moran J., Herrnstein J., Greenhill L., Nakai N., Diamond P., Inoue M., 1995, Natur, 373, 127. doi:10.1038/373127a0



\bibitem[\protect\citeauthoryear{Narayan \& Yi}{1995}]{1995ApJ...452..710N} Narayan R., Yi I., 1995, ApJ, 452, 710. doi:10.1086/176343

\bibitem[\protect\citeauthoryear{Nemmen, Storchi-Bergmann, \& Eracleous}{2014}]{2014MNRAS.438.2804N} Nemmen R.~S., Storchi-Bergmann T., Eracleous M., 2014, MNRAS, 438, 2804. doi:10.1093/mnras/stt2388

\bibitem[\protect\citeauthoryear{Nied{\'z}wiecki, Xie, \& Zdziarski}{2012}]{2012MNRAS.420.1195N} Nied{\'z}wiecki A., Xie F.-G., Zdziarski A.~A., 2012, MNRAS, 420, 1195. doi:10.1111/j.1365-2966.2011.20106.x

\bibitem[\protect\citeauthoryear{Nied{\'z}wiecki, Xie, \& Stepnik}{2013}]{2013MNRAS.432.1576N} Nied{\'z}wiecki A., Xie F.-G., Stepnik A., 2013, MNRAS, 432, 1576. doi:10.1093/mnras/stt573

\bibitem[\protect\citeauthoryear{Nied{\'z}wiecki, St{\c{e}}pnik, \& Xie}{2015}]{2015ApJ...799..217N} Nied{\'z}wiecki A., St{\c{e}}pnik A., Xie F.-G., 2015, ApJ, 799, 217. doi:10.1088/0004-637X/799/2/217

\bibitem[\protect\citeauthoryear{Nied{\'z}wiecki, Szanecki, \& Zdziarski}{2019}]{2019MNRAS.485.2942N} Nied{\'z}wiecki A., Szanecki M., Zdziarski A.~A., 2019, MNRAS, 485, 2942. doi:10.1093/mnras/stz487

\bibitem[\protect\citeauthoryear{Nied{\'z}wiecki et al.}{2022}]{2022ApJ...931..167N} Nied{\'z}wiecki A., Szanecki M., Zdziarski A.~A., Xie F.-G., 2022, ApJ, 931, 167. doi:10.3847/1538-4357/ac6c8b


\bibitem[\protect\citeauthoryear{Oka \& Manmoto}{2003}]{2003MNRAS.340..543O} Oka K., Manmoto T., 2003, MNRAS, 340, 543. doi:10.1046/j.1365-8711.2003.06476.x


\bibitem[\protect\citeauthoryear{Reid, Pesce, \& Riess}{2019}]{2019ApJ...886L..27R} Reid M.~J., Pesce D.~W., Riess A.~G., 2019, ApJL, 886, L27. doi:10.3847/2041-8213/ab552d

\bibitem[\protect\citeauthoryear{Reynolds et al.}{2009}]{2009ApJ...691.1159R} Reynolds C.~S., Nowak M.~A., Markoff S., Tueller J., Wilms J., Young A.~J., 2009, ApJ, 691, 1159. doi:10.1088/0004-637X/691/2/1159


\bibitem[\protect\citeauthoryear{Schimoia et al.}{2017}]{2017MNRAS.472.2170S} Schimoia J.~S., Storchi-Bergmann T., Winge C., Nemmen R.~S., Eracleous M., 2017, MNRAS, 472, 2170. doi:10.1093/mnras/stx2107

\bibitem[\protect\citeauthoryear{Schnorr-M{\"u}ller et al.}{2014}]{2014MNRAS.438.3322S} Schnorr-M{\"u}ller A., Storchi-Bergmann T., Nagar N.~M., Ferrari F., 2014, MNRAS, 438, 3322. doi:10.1093/mnras/stt2440

\bibitem[\protect\citeauthoryear{Shapiro, Lightman, \& Eardley}{1976}]{1976ApJ...204..187S} Shapiro S.~L., Lightman A.~P., Eardley D.~M., 1976, ApJ, 204, 187. doi:10.1086/154162

\bibitem[\protect\citeauthoryear{Starling et al.}{2005}]{2005MNRAS.356..727S} Starling R.~L.~C., Page M.~J., Branduardi-Raymont G., Breeveld A.~A., Soria R., Wu K., 2005, MNRAS, 356, 727. doi:10.1111/j.1365-2966.2004.08493.x


\bibitem[\protect\citeauthoryear{Ursini et al.}{2015}]{2015MNRAS.452.3266U} Ursini F., Marinucci A., Matt G., Bianchi S., Tortosa A., Stern D., Ar{\'e}valo P., et al., 2015, MNRAS, 452, 3266. doi:10.1093/mnras/stv1527

\bibitem[\protect\citeauthoryear{Wojaczy{\'n}ski et al.}{2015}]{2015A&A...584A..20W} Wojaczy{\'n}ski R., Nied{\'z}wiecki A., Xie F.-G., Szanecki M., 2015, A\&A, 584, A20. doi:10.1051/0004-6361/201526621



\bibitem[\protect\citeauthoryear{Xie et al.}{2016}]{2016MNRAS.463.2287X} Xie F.-G., Zdziarski A.~A., Ma R., Yang Q.-X., 2016, MNRAS, 463, 2287. doi:10.1093/mnras/stw2132

\bibitem[\protect\citeauthoryear{Yamada et al.}{2009}]{2009PASJ...61..309Y} Yamada S., Itoh T., Makishima K., Nakazawa K., 2009, PASJ, 61, 309. doi:10.1093/pasj/61.2.309


\bibitem[\protect\citeauthoryear{Yuan, Quataert, \& Narayan}{2003}]{2003ApJ...598..301Y} Yuan F., Quataert E., Narayan R., 2003, ApJ, 598, 301. doi:10.1086/378716


\bibitem[\protect\citeauthoryear{Yuan \& Narayan}{2014}]{2014ARA&A..52..529Y} Yuan F., Narayan R., 2014, ARA\&A, 52, 529. doi:10.1146/annurev-astro-082812-141003

\bibitem[\protect\citeauthoryear{Zdziarski et al.}{2021}]{2021ApJ...914L...5Z} Zdziarski A.~A., Jourdain E., Lubi{\'n}ski P., Szanecki M., Nied{\'z}wiecki A., Veledina A., Poutanen J., et al., 2021, ApJL, 914, L5. doi:10.3847/2041-8213/ac0147


\end{thebibliography}
\end{document}